\documentclass[11pt]{JHEP3}

\usepackage{amsfonts}
\usepackage{amssymb}
\usepackage{mathrsfs}

\usepackage{amsmath,amssymb}
\usepackage{bbm}

\def\slasha#1{\setbox0=\hbox{$#1$}#1\hskip-\wd0\hbox to\wd0{\hss\sl/\/\hss}}

\def\periodb#1{\setbox0=\hbox{$#1$}#1\hskip-\wd0\hbox to\wd0{-}}
\newcommand{\unit}{\mathbbm{1}}   
\newcommand{\CA}{\mathcal{A}}    
\newcommand{\CAt}{\tilde{\mathcal{A}}}    
\newcommand{\CCA}{\mathscr{A}}    
\newcommand{\CB}{\mathcal{B}}    
\newcommand{\CD}{\mathcal{D}}    
\newcommand{\CF}{\mathcal{F}}    
\newcommand{\CCF}{\mathscr{F}}    
\newcommand{\CCG}{\mathscr{G}}    
\newcommand{\CH}{\mathcal{H}}    
\newcommand{\CK}{\mathcal{K}}    
\newcommand{\CN}{\mathcal{N}}    
\newcommand{\CO}{\mathcal{O}}    
\newcommand{\CP}{\mathcal{P}}    
\newcommand{\CQ}{\mathcal{Q}}    

\newcommand{\CT}{\mathcal{T}}    
\newcommand{\CU}{\mathcal{U}}    
\newcommand{\CV}{\mathcal{V}}    
\newcommand{\CX}{\mathcal{X}}    
\newcommand{\CY}{\mathcal{Y}}    
\newcommand{\CZ}{\mathcal{Z}}    
\newcommand{\CE}{\mathcal{E}}    
\newcommand{\frA}{\mathfrak{A}}    
\newcommand{\FR}{\mathbbm{R}}     
\newcommand{\FC}{\mathbbm{C}}     
\newcommand{\CPP}{{\mathbbm{C}P}}    
\newcommand{\RZ}{\mathbbm{Z}}     
\newcommand{\dd}{\mathrm{d}}     
\newcommand{\dpar}{\partial}     
\newcommand{\dparb}{{\bar{\partial}}}     
\newcommand{\mub}{{\bar{\mu}}}     
\newcommand{\etab}{{\bar{\eta}}}     
\newcommand{\sigmab}{{\bar{\sigma}}}     
\newcommand{\embd}{{\hookrightarrow}}     
\newcommand{\diag}{{\mathrm{diag}}}     
\newcommand{\ed}{{\dot{1}}}    
\newcommand{\zd}{{\dot{2}}}    
\newcommand{\di}{\mathrm{i}}     
\newcommand{\bz}{{\bar{z}}}     
\newcommand{\bl}{{\bar{\lambda}}}     
\newcommand{\hl}{{\hat{\lambda}}}     
\newcommand{\bE}{{\bar{E}}}     
\newcommand{\bV}{{\bar{V}}}     
\newcommand{\ald}{{\dot{\alpha}}}     
\newcommand{\bed}{{\dot{\beta}}}     
\newcommand{\gad}{{\dot{\gamma}}}     
\newcommand{\ded}{{\dot{\delta}}}     
\newcommand{\eps}{{\varepsilon}}     
\newcommand{\eand}{{~~~\mbox{and}~~~}}     
\newcommand{\ewith}{{~~~\mbox{with}~~~}}     
\newcommand{\der}[1]{\frac{\dpar}{\dpar #1}}   
\newcommand{\derr}[2]{\frac{\dpar #1}{\dpar #2}}   
\newcommand{\tr}{\,\mathrm{tr}\,}     

\newcommand{\agl}{\mathrm{gl}}     
\newcommand{\au}{\mathrm{u}}     
\newcommand{\sU}{\mathrm{U}}     
\newcommand{\sSU}{\mathrm{SU}}     
\newcommand{\sSL}{\mathrm{SL}}     
\newcommand{\sGL}{\mathrm{GL}}     
\newcommand{\sSO}{\mathrm{SO}}     
\newcommand{\sSpin}{\mathrm{Spin}}     
\newcommand{\sEnd}{\mathrm{End}\,}     
\newcommand{\sts}{\CP^{3|4}}

\newcommand{\remark}[1]{}     
\newcommand{\z}[1]{{\stackrel{\circ}{#1}}{}}     

\title{Matrix Models and D-Branes \\[0.5cm] in Twistor String Theory}

\author{Olaf Lechtenfeld and Christian S\"{a}mann\\
        Institut f\"{u}r Theoretische Physik\\
        Universit\"{a}t Hannover\\
        Appelstra{\ss}e 2, D-30167 Hannover, Germany\\
        Email: \email{lechtenf@itp.uni-hannover.de}, \email{saemann@itp.uni-hannover.de}
}

\abstract{We construct two matrix models from twistor string
theory: one by dimensional reduction onto a rational curve and
another one by introducing noncommutative coordinates on the
fibres of the supertwistor space $\CP^{3|4}\rightarrow \CPP^1$. We
comment on the interpretation of our matrix models in terms of
topological D-branes and relate them to a recently proposed string
field theory. By extending one of the models, we can carry over
all the ingredients of the super ADHM construction to a D-brane
configuration in the supertwistor space $\CP^{3|4}$. Eventually,
we present the analogue picture for the (super) Nahm
construction.}

\preprint{hep-th/0511130\\ITP--UH--21/05 }


\begin{document}

\section{Introduction}

The basic idea of twistor string theory
\cite{Witten:2003nn}\footnote{See \cite{Berkovits:2004hg} for an
alternative formulation.} is the union of twistor geometry with
Calabi-Yau geometry in the supermanifold $\CPP^{3|4}$. This space
is simultaneously a supertwistor space and a Calabi-Yau
supermanifold and one can use it as a target space for a
topological B-model, which can be shown to be equivalent to
$\CN=4$ supersymmetrically extended self-dual Yang-Mills (SDYM)
theory. By incorporating additional D-instantons into the picture,
one can obtain the full $\CN=4$ supersymmetric Yang-Mills (SYM)
theory and use its twistorial and string theoretical description
for calculating amplitudes in this theory. For a good overview of
the results in this area, see e.g.\ \cite{webpage}.

Already a number of variations and reductions of the underlying
supertwistor space $\CPP^{3|4}$ and its open subset
$\CP^{3|4}=\CPP^{3|4}\backslash \CPP^{1|4}$ have been considered
\cite{Popov:2004nk}-\cite{Lindstrom:2005uh}. In this paper, we
want to discuss dimensional reductions of the bosonic dimensions
of $\CP^{3|4}$ and construct matrix models from the twistor
string. For obtaining these, we will use two methods. Starting
point of both is holomorphic Chern-Simons (hCS) theory on the
noncompact supertwistor space $\CP^{3|4}$, which is a rank $2|4$
(complex) vector bundle over the Riemann sphere $\CPP^1$ and
becomes diffeomorphic to $\FR^{4|8}\times \CPP^1$ after imposing
reality conditions on its sections. Here, $\FR^{4|8}$ is the
moduli space of (real) holomorphic sections of $\CP^{3|4}$.

In the first approach, we will dimensionally reduce $\CP^{3|4}$ to
the rank $0|4$ vector bundle $\CPP^{1|4}$ over $\CPP^1$. Via the
twistor correspondence, this amounts to reducing the moduli space
$\FR^{4|8}$ by its bosonic coordinates to $\FR^{0|8}$:
$\CPP^{1|4}\cong \FR^{0|8}\times \CPP^1$. On the field theory
side, we will obtain an action corresponding to matrix quantum
mechanics with complex time over $\CPP^{1|4}$. For a similar
construction on the conifold, see \cite{Bonelli:2005dc}.

The second method will be to impose a noncommutative algebra on
the bosonic coordinates of the moduli space $\FR^{4|8}$, which
yields a noncommutative algebra for the fibre coordinates of
$\CP^{3|4}$. This turns the derivatives and its coordinates into
operators in an infinite dimensional Fock space, which can be
represented by infinite dimensional matrices. In this sense, hCS
theory will again be reduced to matrix quantum mechanics, as the
integral over the bosonic moduli becomes a trace over the Fock
space.

Starting from hCS theory with gauge group $\sGL(n,\FC)$, the first
method yields a matrix model whose field content takes values in
the Lie algebra of $\sGL(n,\FC)$. One expects this model to be
equivalent to the second one in an appropriate limit $n\rightarrow
\infty$. Furthermore, both models can be reduced by integrating
over the remaining bosonic coordinate of $\CPP^{1|4}$, which leads
to matrix models of $\CN=4$ SDYM theory.

Having defined these matrix models, we will elaborate on their
physical interpretation and discuss their relation to the cubic
string field theory proposed in \cite{Lechtenfeld:2004cc} as well
as their r{\^o}les as effective actions for certain D-brane
configurations.

The D-brane interpretation of the matrix models will always be
twofold: On the one hand, we have physical D-branes in type IIB
superstring theory with the moduli space $\FR^{4|8}$ being a
subspace of the ten-dimensional target space. On the other hand,
we have topological D-branes of B-type topological string theory
in the supertwistor space $\CP^{3|4}$. By extending the matrix
model on $\CPP^{1|4}$, we will be able to carry over all the
ingredients of the D-brane interpretation of the ADHM construction
from the moduli space $\FR^{4|8}$ to the supertwistor space
$\CP^{3|4}$. Introducing further dimensional reductions of hCS
theory on $\CP^{3|4}$, we do the same for the D-brane
configuration describing the super Nahm construction.

Together with \cite{Wimmer:2005bz}, this paper is intended as a
first step towards a D-brane interpretation of solutions to
noncommutative self-dual Yang-Mills theory and dimensional
reductions thereof \cite{Lechtenfeld:ff2}.

The outline of this paper is as follows. Section 2 is devoted to a
review of the geometry of $\CP^{3|4}$, hCS theory on this space
and this theory's relation to $\CN=4$ SDYM theory on $\FR^{4|8}$.
In section 3, we construct the matrix models from both dimensional
reduction and noncommutativity. Furthermore, we point out the
similarity of the matrix models with a string field theory. The
D-brane interpretation of the matrix models and the extension to
an ADHM model in $\CP^{3|4}$ is presented in section 4. The
corresponding picture for the Nahm construction is drawn in
section 5, and we conclude in section 6.

\section{Holomorphic Chern-Simons theory on $\sts$}

Recall that the open topological B-model on a complex
three-dimensional Calabi-Yau manifold with a stack of $n$
D5-branes is equivalent to holomorphic Chern-Simons (hCS) theory
which describes holomorphic structures on a rank $n$ vector bundle
over the same space \cite{Witten:2003nn}. In this section, we will
briefly review the definitions of $\CN=4$ supersymetrically
extended hCS theory on the supertwistor space
$\sts=\CPP^{3|4}\backslash \CPP^{1|4}$ arising from the
topological B-model on $\sts$. For a more detailed discussion, see
e.g.\ \cite{Witten:2003nn,Popov:2004rb}.

\subsection{The complex twistor correspondence}

Consider the Riemann sphere $\CPP^1\cong S^2$ with complex
homogeneous coordinates $\lambda_{\dot{1}}$ and
$\lambda_{\dot{2}}$. We can cover this space by two patches $U_+$
and $U_-$ for which $\lambda_{\dot{1}}\neq 0$ and
$\lambda_{\dot{2}}\neq 0$, respectively, and introduce the
standard (affine) complex coordinates
$\lambda_+:=\frac{\lambda_{\dot{2}}}{\lambda_{\dot{1}}}$ on $U_+$
and $\lambda_-:=\frac{\lambda_{\dot{1}}}{\lambda_{\dot{2}}}$ on
$U_-$ with $\lambda_+=(\lambda_-)^{-1}$ on the overlap $U_+\cap
U_-$.

The sections of the holomorphic line bundle $\CO(1)$ over $\CPP^1$
are described by holomorphic functions $z_\pm$ over $U_\pm$ which
are related by $z_+=\lambda_+ z_-$ on the intersection $U_+\cap
U_-$. Using the parity changing operator $\Pi$, which inverts the
parity of the fibre coordinates when acting on a fibre bundle, we
also define the bundle $\Pi\CO(1)$ whose sections are described by
holomorphic Gra{\ss}mann-valued functions $\eta_\pm$ over $U_\pm$ with
$\eta_+=\lambda_+\eta_-$ on $U_+\cap U_-$.

We can now define the supertwistor space
\begin{equation}
\sts\ :=\ \CPP^{3|4}\backslash\CPP^{1|4} \ =\
\FC^2\otimes\CO(1)\oplus\FC^4\otimes\Pi\CO(1)
\end{equation}
as the total space of a rank $2|4$ holomorphic vector bundle
\begin{equation}\label{2.2}
\sts\ \ \rightarrow\  \ \CPP^1~.
\end{equation}
This bundle can be covered by the two patches
\begin{equation}
\CU_+\ :=\ \left.\sts\right|_{U_+}\ \cong\ U_+\times
\FC^{2|4}_+\eand\CU_-\ :=\ \left.\sts\right|_{U_-}\ \cong\
U_-\times \FC^{2|4}_-
\end{equation}
with coordinates $(z_\pm^\alpha,\lambda_\pm,\eta_i^\pm)$, where
$\lambda_\pm$ are the coordinates on $U_\pm$, $z_\pm^\alpha$ with
$\alpha=1,2$ are the coordinates on the bosonic fibres
$\FC_\pm^{2|0}$ and $\eta_i^\pm$ with $i=1,\ldots,4$ are the
coordinates on the fermionic fibres $\FC^{0|4}_\pm$. For
convenience, we also introduce $z^3_\pm=\lambda_\pm$.

Consider now the complex superspace $\FC^{4|8}$ with
coordinates\footnote{In fact, this space is the (anti-)chiral
superspace with half of the fermionic coordinates of the full
$\CN=4$ superspace $\FC^{4|16}$, see e.g.\ \cite{Popov:2004rb} for
more details.}
\begin{equation}
x^{\alpha\ald}~~~\mbox{on}~~~\FC^{4|0}\eand
\eta_i^\ald~~~\mbox{on}~~~\FC^{0|8}~,~~~\alpha,\ald\ =\ 1,2~,
\end{equation}
and the so-called {\em correspondence space}
\begin{equation}
\CF^{5|8}\ :=\ \FC^{4|8}\times \CPP^1~.
\end{equation}
We can define a projection $\pi_2:\CF^{5|8}\rightarrow \CP^{3|4}$
by the formula
\begin{equation}
\pi_2(x^{\alpha\ald},\lambda_\ald^\pm,\eta_i^\ald)\ :=\
(z^\alpha_\pm,\lambda_\pm,\eta_i^\pm)
\end{equation}
with
\begin{equation}\label{2.6}
z^\alpha_\pm\ :=\ x^{\alpha\ald}\lambda_\ald^\pm\eand \eta_i^\pm\
 :=\ \eta_i^\ald\lambda_\ald^\pm~,
\end{equation}
where
\begin{equation}
(\lambda_\ald^+)\ :=\  \left(\begin{array}{c} 1\\\lambda_+
\end{array}\right)\eand (\lambda_\ald^-)\ :=\ \left(\begin{array}{c}
\lambda_- \\ 1
\end{array}\right)~.
\end{equation}
Note that formul\ae{} \eqref{2.6}, the {\em incidence relations},
with fixed $x^{\alpha\ald},\eta_\ald^i$ define holomorphic
sections of the bundle \eqref{2.2} which are projective lines
$\CPP^1_{x,\eta}\embd\CP^{3|4}$.

There is also the canonical projection $\pi_1:\CF^{5|8}\rightarrow
\FC^{4|8}$ given explicitly by the formula
\begin{equation}\label{2.7}
\pi_1(x^{\alpha\ald},\lambda_\ald^\pm,\eta_i^\ald)\ :=\
(x^{\alpha\ald},\eta_i^\ald)~.
\end{equation}
Together, the two projections $\pi_1$ and $\pi_2$ define the
double fibration
\begin{equation}\label{dblfibrationfourself}
\begin{aligned}
\begin{picture}(50,40)
\put(0.0,0.0){\makebox(0,0)[c]{$\CP^{3|4}$}}
\put(64.0,0.0){\makebox(0,0)[c]{$\FC^{4|8}$}}
\put(34.0,33.0){\makebox(0,0)[c]{$\CF^{5|8}$}}
\put(7.0,18.0){\makebox(0,0)[c]{$\pi_2$}}
\put(55.0,18.0){\makebox(0,0)[c]{$\pi_1$}}
\put(25.0,25.0){\vector(-1,-1){18}}
\put(37.0,25.0){\vector(1,-1){18}}
\end{picture}
\end{aligned}
\end{equation}
which yields a {\em twistor correspondence} between $\CP^{3|4}$
and $\FC^{4|8}$, i.e.\ a correspondence between points in one
space and subspaces of the other one:
\begin{equation*}
\begin{aligned}
\left\{\,\mbox{points $p$ in $\CP^{3|4}$}\right\}\
\longleftrightarrow\ \left\{\,\mbox{null $(2|4)$-dimensional
superplanes $\pi_1(\pi_2^{-1}(p))$ in  $\FC^{4|8}$}\right\}~,\\
\left\{\,\mbox{projective lines $\CPP^1_{x,\eta}\ =\
\pi_2(\pi_1^{-1}(x,\eta))\embd\CP^{3|4}$}\right\}\
\longleftrightarrow\ \left\{\,\mbox{points $(x,\eta)$ in
$\FC^{4|8}$}\right\}~.\hspace{0.25cm}
\end{aligned}
\end{equation*}

\subsection{The real twistor correspondence}

Recall that a real structure on a complex manifold $X$ is defined
as an anti-linear involution $\tau:X\rightarrow X$. On
$\FC^{4|8}$, we can define the following ones:
\begin{equation}
\begin{aligned}
\tau_\eps(x^{1\ed})\ =\ \bar{x}^{2\zd}~,~~~ \tau_\eps(x^{1\zd})\
=\ \eps
\bar{x}^{2\ed}~~~\mbox{with}~~~\eps\ =\ \pm 1~,\hspace{2cm}\\
\tau_{+1}(\eta_i^\ed)\ =\ \etab_i^\zd\eand
\tau_{-1}\left(\begin{array}{cccc} \eta_1^\ed & \eta_2^\ed
&\eta_3^\ed &\eta_4^\ed \\\eta_1^\zd &\eta_2^\zd &\eta_3^\zd
&\eta_4^\zd
\end{array}\right)\ =\ \left(\begin{array}{cccc}
-\bar{\eta}^{\dot{2}}_2 &\bar{\eta}^{\dot{2}}_1
&-\bar{\eta}^{\dot{2}}_4 &\bar{\eta}^{\dot{2}}_3 \\
\bar{\eta}^{\dot{1}}_2 &-\bar{\eta}^{\dot{1}}_1
&\bar{\eta}^{\dot{1}}_4 &-\bar{\eta}^{\dot{1}}_3 \\
\end{array}\right)~.
\end{aligned}
\end{equation}
(A third possible involution $\tau_0$ is given e.g.\ in
\cite{Popov:2004rb}.) The corresponding reality conditions are
then obtained by demanding invariance under the maps $\tau_\eps$,
i.e.\ to impose the conditions
\begin{equation}
x^{1\dot{1}}\ =\ \bar{x}^{2\dot{2}}~,~~~x^{1\dot{2}}\ =\
\eps\bar{x}^{2\dot{1}}~,~~~\left\{
\begin{array}{cl}
\eta_i^{\dot{1}}\ =\ \bar{\eta}_i^{\dot{2}}&\mbox{for }\eps\ =\ +1\\
\begin{array}{cc}
\eta_1^{\dot{1}}\ =\ -\bar{\eta}_2^{\dot{2}}~,&\eta_1^{\dot{2}}\ =\ \bar{\eta}_2^{\dot{1}}\\
\eta_3^{\dot{1}}\ =\ -\bar{\eta}_4^{\dot{2}}~,&\eta_3^{\dot{2}}\ =\ \bar{\eta}_4^{\dot{1}}\\
\end{array}
 & \mbox{for }\eps\ =\ -1~.
\end{array}\right.
\end{equation}
We thus obtain the real superspace $\FR^{4|8}\subset \FC^{4|8}$
together with a natural metric on its body defined by $\dd
s^2=\det (\dd x^{\alpha\ald})$. This metric is of Kleinian
signature $($$-$$-$$+$$+$$)$ for $\eps=+1$ and of Euclidean
signature $($$+$$+$$+$$+$$)$ for $\eps=-1$. The literal
identification with the coordinates $x^\mu$ on $\FR^4$ is then
chosen to be
\begin{equation}\label{2.13}
x^{2\zd}\ =\ \bar{x}^{1\ed}\ =:\ -(\eps x^4+\di x^3)\eand
x^{2\ed}\ =\ \eps\bar{x}^{1\zd}\ =:\ -\eps(x^2-\di x^1)~.
\end{equation}

Defining the anti-holomorphic involution on $\CPP^1$ as
\begin{equation}
\tau_\eps(\lambda_\pm)\ =\ \frac{\eps}{\bl_\pm}
\end{equation}
and using the incidence relations \eqref{2.6}, we obtain
\begin{equation}\label{2.14}
\tau_{\eps}(z_+^1,z_+^2,\lambda_+)\ =\
\left(\frac{\bz_+^2}{\bl_+},
\frac{\eps\bz_+^1}{\bl_+},\frac{\eps}{\bl_+}\right)\eand
\tau_{\eps}(z_-^1,z_-^2,\lambda_-)\ =\
\left(\frac{\eps\bz_-^2}{\bl_-},
\frac{\bz^1_-}{\bl_-},\frac{\eps}{\bl_-}\right)
\end{equation}
on the bosonic coordinates of the supertwistor space $\sts$ and
\begin{equation}\label{2.15}
\tau_1(\eta^\pm_i)\ =\
\left(\frac{\bar{\eta}^\pm_i}{\bl_\pm}\right)~,~~~
\tau_{-1}(\eta^\pm_1,\eta^\pm_2,\eta^\pm_3,\eta^\pm_4)\ =\
\left(\frac{\mp\bar{\eta}^\pm_2}{\bl_\pm},\frac{\pm\bar{\eta}^\pm_1}{\bl_\pm},
\frac{\mp\bar{\eta}^\pm_4}{\bl_\pm},\frac{\pm\bar{\eta}^\pm_3}{\bl_\pm}\right)
\end{equation}
on its fermionic coordinates. It is obvious from \eqref{2.14} and
\eqref{2.15} that the involution $\tau_{-1}$ has no fixed points,
but does leave invariant projective lines $\CPP^1_{x,\eta}\embd
\CP^{3|4}$ with $(x,\eta)\in \FR^{4|8}$. On the other hand, the
involution $\tau_1$ does have fixed points which form a real
supermanifold $\CT^{3|4}$ with coordinates
$(z^\alpha_\pm,\lambda_\pm,\eta_i^\pm)$ satisfying the reality
conditions
$\tau_1(z_\pm^\alpha,\lambda_\pm,\eta_i^\pm)=(z_\pm^\alpha,\lambda_\pm,\eta_i^\pm)$,
i.e.\
\begin{equation}
z_\pm^2\ =\ \lambda_\pm\bz_\pm^1~,~~~\lambda_\pm\bl_\pm\ =\
1~,~~~\eta_i^\pm\ =\ \lambda_\pm\etab_i^\pm~.
\end{equation}

For the space $\FR^{4|8}$, we introduce a new correspondence space
$\FR^{4|8}\times \CPP^1$ with the same projections \eqref{2.6}
onto $\CP^{3|4}$ and \eqref{2.7} onto $\FR^{4|8}$ as well as the
double fibration
\begin{equation}\label{2.17}
\begin{aligned}
\begin{picture}(50,40)
\put(0.0,0.0){\makebox(0,0)[c]{$\CP^{3|4}$}}
\put(64.0,0.0){\makebox(0,0)[c]{$\FR^{4|8}$}}
\put(34.0,33.0){\makebox(0,0)[c]{$\FR^{4|8}\times \CPP^1$}}
\put(7.0,18.0){\makebox(0,0)[c]{$\pi_2$}}
\put(55.0,18.0){\makebox(0,0)[c]{$\pi_1$}}
\put(25.0,25.0){\vector(-1,-1){18}}
\put(37.0,25.0){\vector(1,-1){18}}
\end{picture}
\end{aligned}
\end{equation}
This diagram describes very different situations in the Euclidean
and the Kleinian case. For $\eps=-1$, the map $\pi_2$ is a
diffeomorphism,
\begin{equation}
\CP^{3|4}\ \cong\  \FR^{4|8}\times \CPP^1~,
\end{equation}
and the double fibration \eqref{2.17} is simplified to the
non-holomorphic fibration
\begin{equation}\label{2.19}
\pi_1:~\CP^{3|4}\ \rightarrow\  \FR^{4|8}
\end{equation}
where $3|4$ stands for complex and $4|8$ for real dimensions.
Correspondingly, we can choose either coordinates
$(z^\alpha_\pm,z_\pm^3:=\lambda_\pm,\eta_i^\pm)$ or
$(x^{\alpha\ald},\lambda_\pm,\eta_i^\pm)$ on $\CP^{3|4}$ and
consider this space as a complex $3|4$-dimensional or a real
$6|8$-dimensional manifold.

In the case of Kleinian signature $($$-$$-$$+$$+$$)$, we have
local isomorphisms
\begin{equation}
\sSO(2,2)\ \cong\  \sSpin(2,2)\ \cong\  \sSL(2,\FR)\times
\sSL(2,\FR)\ \cong\ \sSU(1,1)\times \sSU(1,1)
\end{equation}
and under the action of the group $\sSU(1,1)$, the Riemann sphere
$\CPP^1$ of projective spinors decomposes into the disjoint union
$\CPP^1=H^2_+\,\cup\, S^1\,\cup\, H_-^2=H^2\,\cup\, S^1$ of three
orbits. Here, $H^2=H_+^2\cup H_-^2$ is the two-sheeted hyperboloid
and $H_\pm^2=\{\lambda_\pm\in U_\pm\,|\,|\lambda_\pm|<1\}\cong
\sSU(1,1)/\sU(1)$ are open discs. This induces a decomposition of
the correspondence space into
\begin{equation}
\FR^{4|8}\times \CPP^1\ =\ \FR^{4|8}\times H_+^2\cup
\FR^{4|8}\times S^1\cup \FR^{4|8}\times H_-^2\ =\ \FR^{4|8}\times
H^2\cup \FR^{4|8}\times S^1
\end{equation}
as well as a decomposition of the twistor space
\begin{equation}
\CP^{3|4}\ =\ \CP_+^{3|4}\cup \CP_0\cup\CP_-^{3|4}\ =:\
\tilde{\CP}^{3|4}\cup\CP_0~,
\end{equation}
where $\CP_\pm^{3|4}:=\CP^{3|4}|_{H^2_\pm}$ are restrictions of
the rank $2|4$ holomorphic vector bundle \eqref{2.2} to bundles
over $H^2_\pm$. The space $\CP_0:=\CP^{3|4}|_{S^1}$ is the real
$5|8$-dimensional common boundary of the spaces $\CP^{3|4}_\pm$.
There is a natural map from $\FR^{4|8}\times H^2$ into
$\tilde{\CP}^{3|4}$ which is a diffeomorphism between
$\FR^{4|8}\times H_\pm^2$ and $\CP_\pm^{3|4}$,
\begin{equation}\label{2.22}
\tilde{\CP}^{3|4}\ \cong\ \FR^{4|8}\times H^2
\end{equation}
given again by formul\ae{} \eqref{2.6} and their inverses. Thus,
we have a fibration analogously to \eqref{2.19},
\begin{equation}\label{2.23}
\tilde{\CP}^{3|4}\ \rightarrow\ \FR^{4|8}~.
\end{equation}
On the space $\FR^{4|8}\times S^1\subset \FR^{4|8}\times \CPP^1$,
the map \eqref{2.6} becomes a real fibration
\begin{equation}
\FR^{4|8}\times S^1\ \rightarrow\ \CT^{3|4}\ \embd\ \CP_0
\end{equation}
over the real $3|4$-dimensional space. That is why for $(p,q)$
forms, vector fields of type $(0,1)$ etc., we should consider the
spaces \eqref{2.22} and the fibration \eqref{2.23}. However,
holomorphic vector bundles which are described by solutions of hCS
theory on $\tilde{\CP}^{3|4}\subset \CP^{3|4}$ can be extended to
bundles over the whole twistor space. For more details on this,
see e.g.\ \cite{Popov:2004rb,Popov:2005uv}. To indicate which
spaces we are working with, we will use the notation
$\CP^{3|4}_{\eps}$ and imply $\CP^{3|4}_{-1}:=\CP^{3|4}$ and
$\CP^{3|4}_{+1}:=\tilde{\CP}^{3|4}\subset \CP^{3|4}$.

\subsection{Antiholomorphic vector fields on $\sts_\eps$}

On $\CP_\eps^{3|4}$, there is the following relationship between
vector fields of type (0,1) in the coordinates
$(z_\pm^\alpha,z_\pm^3,\eta_i^\pm)$ and vector fields in the
coordinates $(x^{\alpha\ald},\lambda_\pm,\eta_i^\ald)$:
\begin{equation}\label{eq:2.21}
\begin{aligned}
\der{\bz_\pm^1}&\ =\ -\gamma_\pm\lambda_\pm^\ald\der{x^{2\ald}} \
=:\ -\gamma_\pm\bar{V}_2^\pm~,~~~& \der{\bz_\pm^2}&\ =\
\gamma_\pm\lambda_\pm^\ald\der{x^{1\ald}} \ =:\
-\eps\gamma_\pm\bar{V}_1^\pm~,\\\der{\bz_+^3}&\ =\
\der{\bl_+}+\eps\gamma_+
x^{\alpha\dot{1}}\bar{V}_\alpha^++\eps\gamma_+\eta_i^\ed\bV^i_+~,~&
\der{\bz_-^3}&\ =\ \der{\bl_-}+\gamma_-
x^{\alpha\dot{2}}\bar{V}_\alpha^-+\gamma_-\eta_i^\zd\bV^i_-~.
\end{aligned}
\end{equation}
In the Kleinian case, one obtains for the fermionic vector fields
\begin{equation}
\der{\etab^\pm_i}\ =\ -\gamma_\pm \bV^i_\pm\ :=\
-\gamma_\pm\lambda^\ald_\pm \der{\eta^\ald_i}~,
\end{equation}
while in the Euclidean case, we have
\begin{equation}\label{eq:2.37}
\begin{aligned}
\der{\etab_1^\pm}&\ =\
\gamma_\pm\lambda^\ald_\pm\der{\eta_2^\ald}\ =:\
\gamma_\pm\bV_\pm^2~,& \der{\etab_2^\pm}&\ =\
-\gamma_\pm\lambda^\ald_\pm\der{\eta_1^\ald}\ =:\
-\gamma_\pm\bV_\pm^1~,\\ \der{\etab_3^\pm}&\ =\
\gamma_\pm\lambda^\ald_\pm\der{\eta_4^\ald}\ =:\
\gamma_\pm\bV_\pm^4~,& \der{\etab_4^\pm}&\ =\
-\gamma_\pm\lambda^\ald_\pm\der{\eta_3^\ald}\ =:\
-\gamma_\pm\bV_\pm^3~.
\end{aligned}
\end{equation}
All these relations follow from the formul\ae{} \eqref{2.6} and
their inverses. In the above equations, we introduced the factors
\begin{equation}\label{gammas}
\gamma_+\ =\ \frac{1}{1-\eps\lambda_+\bl_+}\ =\ \frac{1}{\hl^\ald_+\lambda^+_\ald}\eand
\gamma_-\ =\ -\eps\frac{1}{1-\eps\lambda_-\bl_-}\ =\ \frac{1}{\hl^\ald_-
\lambda^-_\ald}~,
\end{equation}
where indices are raised and lowered as usual with the
antisymmetric tensor of $\sSL(2,\FC)$. For the latter, we use the
convention $\eps^{\ed\zd}=-\eps_{\ed\zd}=1$ (which implies that
$\eps_{\ald\bed}\eps^{\bed\gad}=\delta_\ald^\gad$). The
coordinates $\hl^\pm_\ald$ are obtained from the coordinates
$\lambda^\pm_\ald$ by an appropriate action of the real structure
$\tau_\eps$ \cite{Popov:2004rb}. To obtain the coordinates
$\hl_\pm^\ald$, one first raises the index and then applies the
action of $\tau$. Altogether, we have the following variants of
the two-spinor $\lambda_\ald^\pm$:
\begin{align}\label{alllambdas}
(\lambda_+^\ald)\ :=\ &(\eps^{\ald\bed}\lambda^+_\bed)\ =\
\left(\begin{array}{c} \lambda_+\\-1
\end{array}\right)~,~~~(\lambda_-^\ald)\ :=\ \left(\begin{array}{c}
1\\ -\lambda_-
\end{array}\right)~,\\\nonumber
(\hl^+_\ald)\ :=\ \left(\begin{array}{c} \eps\bl_+\\1
\end{array}\right)~,~~
(\hl^-_\ald)&\ :=\ \left(\begin{array}{c} \eps\\\bl_-
\end{array}\right)~,~~(\hl_+^\ald)\ :=\ \left(\begin{array}{c}
-\eps\\ \bl_+
\end{array}\right)~,~~
(\hl_-^\ald)\ :=\ \left(\begin{array}{c} -\eps\bl_-\\1
\end{array}\right)~.
\end{align}

\subsection{Forms on $\sts_\eps$}

One can introduce the (nowhere vanishing) holomorphic volume form
\begin{equation}\label{Omega}
\Omega_\pm\ :=\ \left.{\Omega}\right|_{\CU_\pm}\ :=\ \pm\dd
\lambda_\pm\wedge\dd z^1_\pm\wedge \dd z^2_\pm \dd
\eta^\pm_1\ldots\dd\eta^\pm_4\ =:\ \pm\dd \lambda_\pm\wedge\dd
z^1_\pm\wedge \dd z^2_\pm~\Omega^\eta_\pm
\end{equation}
on $\CP^{3|4}$. The existence of this volume element implies that
the Berezinian line bundle is trivial and consequently $\CP^{3|4}$
is a Calabi-Yau supermanifold \cite{Witten:2003nn}. Note, however,
that $\Omega$ is not a differential form because its fermionic
part transforms as a product of Gra{\ss}mann-odd vector fields, i.e.\
with the inverse of the Jacobian. Such forms are called integral
forms.

It will also be useful to introduce $(0,1)$-forms $\bE^a_\pm$ and
$\bE_i^\pm$ which are dual to $\bV_a^\pm$ and $\bV^i_\pm$,
respectively, i.e.\
\begin{equation}
\bV_a^\pm \lrcorner\bE^b_\pm\ =\ \delta_a^b\eand
\bV^i_\pm\lrcorner \bE_j^\pm\ =\ \delta_j^i~.
\end{equation}
Here, $\lrcorner$ denotes the interior product of vector fields
with differential forms. Explicitly, the dual $(0,1)$-forms are
given by the formul\ae{}
\begin{equation}\label{DefForms}
\bE^\alpha_\pm\ =\ -\gamma_\pm\hl_\ald^\pm\dd x^{\alpha\ald}~,~~~
\bE^3_\pm\ =\ \dd \bl_\pm\eand\bE_i^\pm\ =\ -\gamma_\pm\hl_\ald^\pm\dd
\eta^\ald_i~.
\end{equation}

\subsection{Holomorphically embedded submanifolds and their normal bundles}

Equations \eqref{2.6} describe a holomorphic embedding of the
space $\CPP^1$ into the supertwistor space $\sts$. That is, for
fixed moduli $x^{\alpha\ald}$ and $\eta^\ald_i$, equations
\eqref{2.6} yield a projective line $\CPP^1_{x,\eta}$ inside the
supertwistor space. The normal bundle to any $\CPP_{x,\eta}^1\embd
\CP^{3|4}$ is
$\CN^{2|4}=\FC^2\otimes\CO(1)\oplus\FC^4\otimes\Pi\CO(1)$ and we
have
\begin{equation}
h^0(\CPP_{x,\eta}^1,\CN^{2|4})\ =\ \dim_\FC
H^0(\CPP^1_{x,\eta},\CN^{2|4})\ \ =\ 4|8~.
\end{equation}
Furthermore, there are no obstructions to the deformation of the
$\CPP^{1|0}_{x,\eta}$ inside $\CP^{3|4}$ since
$h^1(\CPP^1_{x,\eta},\CN^{2|4})=0|0$.

On the other hand, one can fix only the even moduli
$x^{\alpha\ald}$ and consider a holomorphic embedding
$\CPP^{1|4}_x\embd\sts$ defined by the equations
\begin{equation}\label{moduli2}
z^\alpha_\pm\ =\ x^{\alpha\ald}\lambda_\ald^\pm~.
\end{equation}
Recall that the normal bundle to $\CPP^{1|0}_x\embd \CP^{3|0}$ is
the rank two vector bundle $\CO(1)\oplus\CO(1)$. In the supercase,
the formal definition of the normal bundle by the short exact
sequence
\begin{equation}\label{normalbundle14}
0\ \rightarrow\ T\CPP^{1|4}\ \rightarrow\ T\CP^{3|4}|_{\CPP^{1|4}}
\ \rightarrow\ \CN^{2|0}\ \rightarrow\ 0
\end{equation}
yields that $\CN^{2|0}=T\CP^{3|4}|_{\CPP^{1|4}}/T\CPP^{1|4}$ is a
rank two holomorphic vector bundle over $\CPP^{1|4}$ which is (in
the real case) locally spanned by the vector fields $\gamma_\pm
V^\pm_\alpha$, where $V^\pm_\alpha$ is the complex conjugate of
$\bV^\pm_\alpha$. A global section of $\CN^{2|0}$ over
$\CU_\pm\cap \CPP^{1|4}$ is of the form
$s_\pm=T_\pm^\alpha\gamma_\pm V_\alpha^\pm$. Obviously, the
transformation of the components $T^\alpha_\pm$ from patch to
patch is given by $T^\alpha_+=\lambda_+T^\alpha_-$, i.e.\
$\CN^{2|0}=\CO(1)\oplus\CO(1)$.

\subsection{Holomorphic Chern-Simons theory}

Consider a trivial rank $n$ complex vector bundle $\CE$ over
$\sts_\eps$ and a connection one-form $\CA$ on $\CE$. We define
hCS theory by the action
\begin{equation}\label{actionhCS}
S_{\mathrm{hCS}}\ :=\
\int_{\CZ_\eps}\Omega\wedge\tr\left(\CA^{0,1}\wedge\dparb\CA^{0,1}+
\tfrac{2}{3}\CA^{0,1}\wedge\CA^{0,1}\wedge\CA^{0,1}\right)~,
\end{equation}
where $\CA^{0,1}$ is the $(0,1)$-part of $\CA$ which we assume to
satisfy the conditions $\bV_\pm^i\lrcorner \CA^{0,1}=0$ and
$\bV^i_\pm(\bV_a^\pm\lrcorner \CA^{0,1})=0$ for $a=1,2,3$.
Furthermore, $\Omega$ is the holomorphic volume form \eqref{Omega}
and $\CZ_\eps$ is the subspace of $\sts_\eps$ for
which\footnote{This condition is not a contradiction to
$\eta_i^\pm\neq 0$, but merely a restriction of all functions on
$\sts_\eps$ to be holomorphic in the $\eta_i^\pm$.}
$\bar{\eta}_i^\pm=0$ \cite{Witten:2003nn}. The trace is taken over
the gauge group $\sGL(n,\FC)$.

The equations of motion for \eqref{actionhCS} read
\begin{equation}\label{eomhCS}
\dparb\CA^{0,1}+\CA^{0,1}\wedge \CA^{0,1}\ =\ 0~.
\end{equation}
In the following, we will usually discuss them using the
components
\begin{equation}
\CA^\pm_\alpha\ :=\ \bV^\pm_\alpha\lrcorner\,\CA^{0,1}~,~~~
\CA_{\bl_\pm}\ :=\ \der{\bl_\pm}\lrcorner\,\CA^{0,1}~,~~~
\CA^{i}_\pm\ :=\ \bV^i_\pm\lrcorner\,\CA^{0,1}~.
\end{equation}
in which \eqref{eomhCS} takes, e.g.\ on $\CU_+$, the form
\begin{subequations}\label{eomhCScomp}
\begin{eqnarray}\label{shCS1}
\bV_\alpha^+\CA_\beta^+- \bV_\beta^+\CA_\alpha^++
[\CA_\alpha^+,\CA_\beta^+]&=&0~,\\
\label{shCS2} \dpar_{\bl_+}\CA_\alpha^+- \bV_\alpha^+\CA_{\bl_+}+
[\CA_{\bl_+},\CA_\alpha^+]&=&0~.
\end{eqnarray}
\end{subequations}
Using these components, we can rewrite the action
\eqref{actionhCS} as
\begin{equation}\label{actionhCScpt}
S_{\mathrm{hCS}}\ :=\ \int_{\CZ_\eps}\dd \lambda \wedge\dd \bl
\wedge\dd z^1\wedge \dd z^2\wedge E^1\wedge E^2~\Omega^\eta \tr
\eps^{abc}\left(\CA_a \bV_b\CA_c+
\tfrac{2}{3}\CA_a\CA_b\CA_c\right)~.
\end{equation}

Recall that we assumed in \eqref{actionhCS} that $\CA^i_\pm=0$.
Moreover, note that in the twistor approach, one considers those
gauge potentials $\CA^{0,1}$ for which the components
$\CA_{\bl_\pm}$ can be gauged away \cite{Popov:2004rb} and thus
restricts the solution space to \eqref{eomhCS} to a
subset\footnote{This subset contains in particular the vacuum
solution $\CA^{0,1}=0$ and its vicinity.}. In this case, one can
choose a gauge such that the superfield expansion of
$\CA_\alpha^\pm$ and $\CA_{\bl_\pm}$ in $\eta_i^\pm$ and
$\lambda_\pm$ are given by the formul\ae{} \cite{Popov:2004rb}
\begin{subequations}\label{fieldexpansion}
\begin{eqnarray}\label{expAa}
\CA_\alpha^+&=&\lambda_+^\ald\,
A_{\alpha\ald}(x)+\eta_i^+\chi^i_\alpha(x)+
\gamma_+\,\tfrac{1}{2!}\,\eta^+_i\eta^+_j\,\hl^\ald_+\,
\phi_{\alpha \ald}^{ij}(x)+\\
\nonumber
&&+\gamma_+^2\,\,\tfrac{1}{3!}\,\eta^+_i\eta^+_j\eta^+_k\,\hl_+^\ald\,
\hl_+^\bed\,
\tilde{\chi}^{ijk}_{\alpha\ald\bed}(x)+\gamma_+^3\,\tfrac{1}{4!}\,
\eta^+_i\eta^+_j\eta^+_k\eta^+_l\,
\hl_+^\ald\,\hl_+^\bed\,\hl_+^{\dot{\gamma}}\,
G^{ijkl}_{\alpha\ald\bed\dot{\gamma}}(x)~,\\
\label{expAl}
\CA_{\bl_+}&=&\gamma_+^2\eta^+_i\eta^+_j\,\phi^{ij}(x)-
\gamma_+^3\eta^+_i\eta^+_j\eta^+_k\,\hl_+^\ald\,
\tilde{\chi}^{ijk}_{\ald}
(x)+\\
\nonumber &&+2\gamma_+^4\eta^+_i\eta^+_j\eta^+_k\eta^+_l\,
\hl_+^\ald\,\hl_+^\bed G^{ijkl}_{\ald\bed}(x)~.
\end{eqnarray}
\end{subequations}
Together with these expansions, the field equations
\eqref{eomhCScomp} of hCS theory on $\CZ_\eps$ are reduced to the
equations of motion of $\CN=4$ supersymmetric self-dual Yang-Mills
(SDYM) theory on $\FR^4_\eps=(\FR^4,g_\eps)$ where
$g_{-1}=\diag(+1,+1,+1,+1)$ and $g_{+1}=\diag(-1,-1,+1,+1)$
\cite{Witten:2003nn,Popov:2004rb}.

Besides the gauge chosen above, there is also a gauge in which
$\CA_{\bl_\pm}=0$ and $\CA_\pm^i\neq 0$ \cite{Popov:2004rb} and
after performing the (super-)gauge transformation
\begin{equation}\label{gaugetrafo}
(\CA_\alpha^\pm\neq 0,\CA_{\bl_\pm}\neq 0,\CA^i_\pm=0)\
\stackrel{\varphi}{\longrightarrow}\ (\CAt_\alpha^\pm\neq
0,\CAt_{\bl_\pm} \ =\  0,\CAt^i_\pm\neq 0)~,
\end{equation}
the hCS field equations \eqref{eomhCScomp} are transformed to
\begin{subequations}\label{eomhCShat}
\begin{align}
\bV_\alpha^\pm\CAt_\beta^\pm-\bV_\beta^\pm\CAt_\alpha^\pm+[\CAt_\alpha^\pm,\CAt_\beta^\pm]&\
=\ 0~,\\
\bV^i_\pm\CAt_\alpha^\pm+\bV_\alpha^\pm\CAt^i_\pm+\{\CAt^i_\pm,\CAt_\alpha^\pm\}&\
=\ 0~,\\
\bV^i_\pm\CAt^j_\pm+\bV^j_\pm\CAt^i_\pm+\{\CAt^i_\pm,\CAt^j_\pm\}&\
=\ 0~,\\
\label{eomhCShat4} \dpar_{\bl_\pm}\CAt_\alpha^\pm&\ =\ 0~,\\
\dpar_{\bl_\pm}\CAt_\pm^i&\ =\ 0~.
\end{align}
\end{subequations}
Note, however, that these equations cannot be obtained from an
action principle. From \eqref{eomhCShat4} and a generalized
Liouville theorem, it follows that the non-trivial components of
the gauge potential are linear in $\lambda_\pm$ and one can
therefore write
\begin{equation}\label{decomposeA}
\CAt_\alpha^+\ =\ \lambda^\ald_\pm\CAt_{\alpha\ald}\eand
\CAt^i_\pm\
 =\ \lambda^\ald_\pm\CAt^i_\ald~.
\end{equation}
Substituting \eqref{decomposeA} into \eqref{eomhCShat}, we arrive
at the equivalent equations
\begin{equation}\label{SDYMconstraints}
[\nabla_{\alpha\ald},\nabla_{\beta\bed}]\ =\ \eps_{\ald\bed}
\CF_{\alpha\beta},~~~ [\nabla^i_{\ald},\nabla_{\beta\bed}]\ =\
\eps_{\ald\bed} \CF^i_{\beta},~~~
\{\nabla^i_{\ald},\nabla^j_{\bed}\}\ =\ \eps_{\ald\bed} \CF^{ij}~,
\end{equation}
where $\CF_{\alpha\beta}$ and $\CF^{ij}$ are symmetric and
antisymmetric, respectively, in their two indices. Here, we have
introduced the differential operators
\begin{equation}\label{eompotentials}
\nabla_{\alpha\ald}\ :=\ \dpar_{\alpha\ald}+\CAt_{\alpha\ald}\eand
\nabla^i_\ald\ :=\ \dpar_\ald^i+\CAt^i_\ald~,
\end{equation}
together with the self-dual super field strength with components
$(\CF_{\alpha\beta},\CF_\beta^i,\CF^{ij})$ for the super gauge
potential $(\CAt_{\alpha\ald},\CAt_\ald^i)$. The equations
\eqref{SDYMconstraints} are equivalent to \eqref{eomhCScomp}, and
the expansions in the fermionic coordinates $\eta_i^\ald$ of the
components of the super gauge potential and the super field
strength contain all the physical fields.

For convenience, we will impose additionally a transverse gauge
condition \cite{Harnad:1984vk}, and demand that
\begin{equation}\label{transversegauge}
\eta^\ald_i\CAt^i_\ald\ =\ 0~.
\end{equation}
The residual gauge symmetry is the ordinary gauge symmetry of
supersymmetric SDYM theory. In this transverse gauge, the
expansions of the gauge potential and the corresponding field
strength in the fermionic coordinates are well known, see e.g.\
\cite{Devchand:1996gv}. It is usually sufficient to know that
\begin{align}\label{fieldexpansiondevchand}
\CAt^\pm_\alpha&\ =\ \lambda^\ald_\pm\CAt_{\alpha\ald}\ =\
\lambda^\ald_\pm\left(A_{\alpha\ald}-
\eps_{\ald\bed}\eta^\bed_i\chi^i_\alpha+\ldots -\tfrac{1}{12}
\eps_{\ald\bed}\eta^\bed_i\eta^\gad_j\eta^\ded_k\eta^{\dot{\eps}}_l\nabla_{\alpha\gad}
G_{\ded\dot{\eps}}^{ijkl}~\right)~,\\ \nonumber \CAt^i_\pm&\ =\
\lambda^\ald_\pm\CAt^i_\ald\ =\ \lambda^\ald_\pm\left(
\eps_{\ald\bed}\eta^\bed_j\phi^{ij}+\tfrac{2}{3}\eps_{\ald\bed}\eta^\bed_j\eta^\gad_k\tilde{\chi}^{ijk}_\gad
+\tfrac{1}{4}\eps_{\ald\bed}\eta^\bed_j\eta^\gad_k\eta^\ded_l
\left(G^{ijkl}_{\gad\ded}+\eps_{\gad\ded}\ldots \right)\right)~,
\end{align}
as this already determines the other terms in the expansion
completely. The equations \eqref{SDYMconstraints} are satisfied if
and only if the $\CN=4$ supersymmetric SDYM equations on
$(\FR^4,g_\eps)$ hold for the physical fields appearing in the
expansion \eqref{fieldexpansiondevchand}.

\subsection{Supersymmetric self-dual Yang-Mills theory}

The field content of $\CN=4$ supersymmetric SDYM theory is given
by the supermultiplet $(f_{\alpha\beta},\chi^i_\alpha,
\phi^{ij},\tilde{\chi}^{ijk}_{\ald},G^{ijkl}_{\ald\bed})$, whose
components have helicities $(+1,+\frac{1}{2},0,-\frac{1}{2},-1)$,
respectively. They are combined into the action
\cite{Siegel:1992za}
\begin{equation}\label{SDYMaction}
S_{\mathrm{SDYM}}\ =\ \int\! \dd^4 x\,\tr
\left(G^{\ald\bed}f_{\ald\bed}+\tfrac{\eps}{2}\eps_{ijkl}\tilde{\chi}^{\ald
ijk} \nabla_{\alpha\ald}\chi^{\alpha
l}+\tfrac{\eps}{2}\eps_{ijkl}\phi^{ij}\square\phi^{kl}+\eps_{ijkl}\phi^{ij}\chi^{\alpha
k}\chi_\alpha^l\right),
\end{equation}
where we introduced the shorthand notations
$\square:=\frac{1}{2}\nabla_{\alpha\ald}\nabla^{\alpha\ald}$ and
$G^{\ald\bed}:=\frac{1}{2}\eps_{ijkl}G^{\ald\bed ijkl}$. Note that
the trace now is taken over the Lie algebra $\au(n)$. This action
can also be obtained by substituting \eqref{fieldexpansion} into
\eqref{actionhCS} and integrating over $\eta_i^\pm$ and
$\lambda_\pm$, $\bl_\pm$. The corresponding equations of
motion\footnote{Recall that these equations are equivalent to both
the holomorphic Chern-Simons equations \eqref{eomhCScomp} with the
expansion \eqref{fieldexpansion} and the equations
\eqref{eomhCShat} with the expansion
\eqref{fieldexpansiondevchand}.} read
\begin{equation}\label{SDYMeom}
\begin{aligned}
f_{\ald\bed}&\ =\ 0~,\\
\nabla_{\alpha\ald}\chi^{\alpha i}&\ =\ 0~,\\
\Box\phi^{ij}&\ =\ -\tfrac{\eps}{2}\{\chi^{\alpha i},\chi^j_{\alpha}\}~,\\
\nabla_{\alpha\ald}\tilde{\chi}^{\ald ijk}&\ =\ +2
\eps\,[\phi^{[ij},\chi^{k]}_{\alpha}]~,\\
\eps^{\ald\dot{\gamma}}\nabla_{\alpha\ald}G^{ijkl}_{\dot{\gamma}
\dot{\delta}}&\ =\
+\eps\{\chi^{[i}_{\alpha},\tilde{\chi}^{jkl]}_{\dot{\delta}}\}
-\eps\,[\phi^{[ij},\nabla_{\alpha\dot{\delta}}\phi^{kl]}]~,
\end{aligned}
\end{equation}
where we used the following decomposition of the field strength in
self-dual and anti-self-dual parts:
\begin{equation}
F_{\alpha\ald\beta\bed}\ :=\ [\nabla_{\alpha\ald},\nabla_{\beta\bed}]\ =\ \eps_{\ald\bed}f_{\alpha\beta}+
\eps_{\alpha\beta}f_{\ald\bed}~.
\end{equation}

The supersymmetric SDYM equations for $\CN<4$ are obtained by
considering the first $\CN+1$ equations of \eqref{SDYMeom}. In
addition, one has to restrict the R-symmetry indices $i,j,\ldots$
of all the fields to run from $1$ to $\CN$. One should stress that
for $\CN<4$, the supersymmetric SDYM equations describe indeed a
subsector of the corresponding full SYM theory. For $\CN=4$ the
field contents of both the self-dual and the full theory are
identical, but the interactions differ.

\section{Matrix models}

In this section, we construct four different matrix models. We
start with dimensionally reducing $\CN=4$ SDYM theory to a point,
which yields the first matrix model. The matrices here are just
finite-dimensional matrices from the Lie algebra of the gauge
group $\sU(n)$. The second matrix model we consider results from a
dimensional reduction of hCS theory on $\CP^{3|4}_\eps$ to a
subspace $\CP_\eps^{1|4}\subset\CP^{3|4}_\eps$. We obtain a form
of matrix quantum mechanics with a complex ``time''. This matrix
model is linked by a Penrose-Ward transform to the first matrix
model.

By considering again $\CN=4$ SDYM theory, but on noncommutative
spacetime, we obtain a third matrix model. Here, we have
finite-dimensional matrices with operator entries which can be
realized as infinite-dimensional matrices acting on the tensor
product of the gauge algebra representation space and the Fock
space. The fourth and last matrix model is obtained by rendering
the fibre coordinates in the vector bundle
$\CP^{3|4}_\eps\rightarrow \CPP^{1|4}$ noncommutative. In the
operator formulation, this again yields a matrix model with
infinite-dimensional matrices and there is also a Penrose-Ward
transform which renders the two noncommutative matrix models
equivalent.

In a certain limit, in which the rank $n$ of the gauge groups
$\sU(n)$ and $\sGL(n,\FC)$ of the SDYM and the hCS matrix model
tends to infinity, one expects them to become equivalent to the
respective matrix models obtained from noncommutativity.

\subsection{Matrix model of $\CN=4$ SDYM theory}

We start from the Lagrangian in the action \eqref{SDYMaction} of
$\CN=4$ supersymmetric self-dual Yang-Mills theory in four
dimensions with gauge group $\sU(n)$. One can dimensionally reduce
this theory to a point by assuming that all the fields are
independent of $x\in\FR^4$. This yields the matrix model action
\begin{equation}\label{SDYMactionMM}
\begin{aligned}
S_{\mathrm{SDYMMM}}\ =\
\tr\Bigg(\Bigg.&G^{\ald\bed}\left(-\tfrac{1}{2}\eps^{\alpha\beta}[A_{\alpha\ald},A_{\beta\bed}]\right)+
\tfrac{\eps}{2}\eps_{ijkl}\tilde{\chi}^{\ald ijk}
[A_{\alpha\ald},\chi^{\alpha l}]\\
&+\Bigg. \tfrac{\eps}{4}\eps_{ijkl}
\phi^{ij}[A_{\alpha\ald},[A^{\alpha\ald},\phi^{kl}]]+
\eps_{ijkl}\phi^{ij}\chi^{\alpha k}\chi^l_{\alpha}\Bigg)~,
\end{aligned}
\end{equation}
which is invariant under the adjoint action of the gauge group
$\sU(n)$ on all the fields. This symmetry is the remnant of gauge
invariance. The corresponding equations of motion read
\begin{equation}\label{SDYMeomMM}
\begin{aligned}
\eps^{\alpha\beta}[A_{\alpha\ald},A_{\beta\bed}]&\ =\ 0~,\\
[A_{\alpha\ald},\chi^{\alpha i}]&\ =\ 0~,\\
\tfrac{1}{2}[A^{\alpha\ald},[A_{\alpha\ald},\phi^{ij}]]&\ =\
-\tfrac{\eps}{2}\{\chi^{\alpha
i},\chi^j_{\alpha}\}~,\\
[A_{\alpha\ald},\tilde{\chi}^{\ald ijk}]&\ =\ +
2\eps\,[\phi^{ij},\chi^k_{\alpha},]~,\\
\eps^{\ald\dot{\gamma}}[A_{\alpha\ald},G^{ijkl}_{\dot{\gamma}
\dot{\delta}}]&\ =\
+\eps\{\chi^{i}_{\alpha},\tilde{\chi}^{jkl}_{\dot{\delta}}\}
-\eps\,[\phi^{ij},[A_{\alpha\dot{\delta}},\phi^{kl}]]~.
\end{aligned}
\end{equation}

Note that these equations can be obtained by dimensionally
reducing equations \eqref{SDYMeom} to a point. On the other hand,
the equations of motion of $\CN=4$ SDYM theory are equivalent to
the constraint equations \eqref{SDYMconstraints} which are defined
on the superspace $\FR^{4|8}$. Therefore, \eqref{SDYMeomMM} are
equivalent to the equations
\begin{equation}\label{SDYMconstrainsMM}
[\CAt_{\alpha\ald},\CAt_{\beta\bed}]\ =\ \eps_{\ald\bed}
\CF_{\alpha\beta}~,~~~ \nabla^i_{\ald}\CAt_{\beta\bed}\ =\
\eps_{\ald\bed} \CF^i_{\beta}~,~~~
\{\nabla^i_{\ald},\nabla^j_{\bed}\}\ =\ \eps_{\ald\bed} \CF^{ij}
\end{equation}
obtained from \eqref{SDYMconstraints} by dimensional reduction to
the supermanifold\footnote{Following the usual nomenclature of
superlines and superplanes, this would be a ``superpoint''.}
$\FR^{0|8}$.

Recall that the IKKT matrix model \cite{Ishibashi:1996xs} can be
obtained by dimensionally reducing $\CN=1$ SYM theory in ten
dimensions or $\CN=4$ SYM in four dimensions to a point. In this
sense, the above matrix model is the self-dual analogue of the
IKKT matrix model.

\subsection{Matrix model from hCS theory}

So far, we have constructed a matrix model for $\CN=4$ SDYM
theory, the latter being defined on the space $(\FR^{4|8},g_\eps)$
with $\eps=-1$ corresponding to Euclidean signature and $\eps=+1$
corresponding to Kleinian signature of the metric on $\FR^4$. The
next step is evidently to ask what theory corresponds to the
matrix model introduced above on the twistor space side.

Recall that for the two signatures on $\FR^4$, we use the
supertwistor spaces
\begin{equation}\label{3.3'}
\CP_\eps^{3|4}\ \cong\  \Sigma^1_\eps \times \FR^{4|8}
\end{equation}
where
\begin{equation}
\Sigma_{-1}^1\ :=\ \CPP^1\eand \Sigma^1_{+1}\ :=\ H^2
\end{equation}
and the two-sheeted hyperboloid $H^2$ is considered as a complex
space. As was discussed in section 3.1, the equations of motion
\eqref{SDYMeomMM} of the matrix model \eqref{SDYMactionMM} can be
obtained from the constraint equations \eqref{SDYMconstraints} by
reducing the space $\FR^{4|8}$ to the supermanifold $\FR^{0|8}$
and expanding the superfields contained in
\eqref{SDYMconstrainsMM} in the Gra{\ss}mann variables $\eta_i^\ald$.
On the twistor space side, this reduction yields the orbit spaces
\begin{equation}\label{3.4'}
\Sigma^1_\eps\times \FR^{0|8}\ =\ \CP^{3|4}_\eps/\CCG~,
\end{equation}
where $\CCG$ is the abelian group of translations generated by the
bosonic vector fields $\der{x^{\alpha\ald}}$. Equivalently, one
can define the spaces $\CP^{1|4}_\eps$ as the orbit spaces
\begin{equation}
\CP^{1|4}_\eps\ :=\ \CP^{3|4}_\eps/\CCG^{1,0}~,
\end{equation}
where $\CCG^{1,0}$ is the complex abelian group generated by the
vector fields $\der{z^\alpha_\pm}$. These spaces with $\eps=\pm 1$
are covered by the two patches $U^\eps_\pm \cong \FC^{1|4}$ and
they are obviously diffeomorphic to the spaces \eqref{3.4'}, i.e.\
\begin{equation}\label{3.5'}
\CP_\eps^{1|4}\ \cong\  \Sigma_\eps^1\times \FR^{0|8}
\end{equation}
due to the diffeomorphism \eqref{3.3'}. In the coordinates
$(z^3_\pm,\eta_i^\pm)$ on $\CP_\eps^{1|4}$ and
$(\lambda_\pm,\eta_i^\ald)$ on $\Sigma^1_\eps\times \FR^{0|8}$,
the diffeomorphism is defined e.g.\ by the formul\ae{}
\begin{equation}\label{eq:2.35}
\begin{aligned}
\eta_1^\ed&\ =\
\frac{\eta_1^+-z_+^3\bar{\eta}_2^+}{1+z_+^3\bz_+^3}\ =\
\frac{\bz_-^3\eta_1^--\bar{\eta}_2^-}{1+z_-^3\bz_-^3}~,&
\eta_2^\ed&\ =\
\frac{\eta_2^++z_+^3\bar{\eta}_1^+}{1+z_+^3\bz_+^3}\ =\
\frac{\bz_-^3\eta_2^-+\bar{\eta}_1^-}{1+z_-^3\bz_-^3}~,\\
\eta_3^\ed&\ =\
\frac{\eta_3^+-z_+^3\bar{\eta}_4^+}{1+z_+^3\bz_+^3}\ =\
\frac{\bz_-^3\eta_3^--\bar{\eta}_4^-}{1+z_-^3\bz_-^3}~,&
\eta_4^\ed&\ =\
\frac{\eta_4^++z_+^3\bar{\eta}_3^+}{1+z_+^3\bz_+^3}\ =\
\frac{\bz_-^3\eta_4^-+\bar{\eta}_3^-}{1+z_-^3\bz_-^3}~,
\end{aligned}
\end{equation}
in the Euclidean case $\eps=-1$. Thus, we have a dimensionally
reduced twistor correspondence between the spaces $\CP_\eps^{1|4}$
and $\FR^{0|8}$
\begin{equation}\label{3.13'}
\begin{aligned}
\begin{picture}(50,40)
\put(0.0,0.0){\makebox(0,0)[c]{$\CP^{1|4}_\eps$}}
\put(64.0,0.0){\makebox(0,0)[c]{$\FR^{0|8}$}}
\put(34.0,33.0){\makebox(0,0)[c]{$\FR^{0|8}\times \Sigma_\eps^1$}}
\put(7.0,18.0){\makebox(0,0)[c]{$\pi_2$}}
\put(55.0,18.0){\makebox(0,0)[c]{$\pi_1$}}
\put(25.0,25.0){\vector(-1,-1){18}}
\put(37.0,25.0){\vector(1,-1){18}}
\end{picture}
\end{aligned}
\end{equation}
where the map $\pi_2$ is the diffeomorphism \eqref{3.5'}. It
follows from \eqref{3.13'} that we have a correspondence between
points $\eta\in \FR^{0|8}$ and subspaces $\CPP^1_\eta$ of
$\CP^{1|4}_\eps$.

Holomorphic Chern-Simons theory on $\CP^{3|4}_\eps$ with the
action \eqref{actionhCS} is defined by a gauge potential
$\CA^{0,1}$ taking values in the Lie algebra of $\sGL(n,\FC)$ and
constrained by the equations $\bV_\pm^i(\bV_a^\pm\lrcorner
\CA^{0,1})=0$, $\bV_\pm^i\lrcorner \CA^{0,1}=0$ for $a=1,2,3$.
After reduction to $\CP_\eps^{1|4}$, $\CA^{0,1}$ splits into a
gauge potential and two complex scalar fields taking values in the
normal bundle $\FC^2\otimes \CO(1)$ to the space
$\CP_\eps^{1|4}\embd \CP_\eps^{3|4}$. In components, we have
\begin{equation}
\CA^{0,1}_{\Sigma\pm}\ =\ \dd \bl_\pm\CA_{\bl_\pm}\eand
\CA_\alpha^\pm\ \rightarrow\
\CX_\alpha^\pm~~~\mbox{on}~~~U_\pm^\eps~,
\end{equation}
where both $\CX_\alpha^\pm$ and $\CA_{\bl_\pm}$ are Lie algebra
valued superfunctions on the subspaces $U_\pm^\eps$ of
$\CP^{1|4}_\eps$. The integral over the chiral subspace
$\CZ_\eps\subset\CP_\eps^{3|4}$ should be evidently substituted by
an integral over the chiral subspace $\CY_\eps\subset
\CP^{1|4}_\eps$. This dimensional reduction of the bosonic
coordinates becomes even clearer with the help of the identity
\begin{equation}\label{formreduction} \dd \lambda_\pm\wedge \dd
\bl_\pm\wedge\dd z_\pm^1\wedge \dd z_\pm^2\wedge \bE_\pm^1\wedge
\bE_\pm^2\ =\ \dd \lambda_\pm\wedge \dd \bl_\pm\wedge\dd
x^{1\dot{1}}\wedge\dd x^{1\dot{2}}\wedge\dd x^{2\dot{1}}\wedge\dd
x^{2\dot{2}}~.
\end{equation}
Altogether, the dimensionally reduced action reads
\begin{equation}\label{actionhCSred}
S_{\mathrm{hCS,red}}\ :=\ \int_{\CY_\eps} \omega \wedge
\tr\eps^{\alpha\beta}\CX_\alpha\left(\dparb\CX_\beta+
\left[\CA_\Sigma^{0,1},\CX_\beta\right]\right)~,
\end{equation}
where the form $\omega$ has components
\begin{equation}\label{defomega}
\omega_\pm\ :=\ \Omega|_{U_\pm^\eps}\ =\ \pm\dd\lambda_\pm
\dd\eta_1^\pm\ldots \dd \eta^\pm_4
\end{equation}
and thus takes values in the bundle $\CO(-2)$. Note furthermore
that $\dparb$ here is the Dolbeault operator on $\Sigma_\eps^1$
and the integral in \eqref{actionhCSred} is well-defined since the
$\CX_\alpha$ take values in the bundles $\CO(1)$. The
corresponding equations of motion are given by
\begin{subequations}\label{shCSred}
\begin{eqnarray}\label{shCSred1}
[\CX_1,\CX_2]&=&0~,\\
\label{shCSred2} \dparb\CX_\alpha+
[\CA^{0,1}_\Sigma,\CX_\alpha]&=&0~.
\end{eqnarray}
\end{subequations}
The gauge symmetry is obviously reduced to the transformations
\begin{equation}
\CX_\alpha\ \rightarrow\ \varphi^{-1}\CX_\alpha\varphi\eand
\CA^{0,1}_\Sigma\ \rightarrow\
\varphi^{-1}\CA^{0,1}_\Sigma\varphi+\varphi^{-1}\dparb\varphi~,
\end{equation}
where $\varphi$ is a smooth $\sGL(n,\FC)$-valued function on
$\CP_\eps^{1|4}$. The matrix model given by \eqref{actionhCSred}
and the field equations \eqref{shCSred} can be understood as
matrix quantum mechanics with a complex ``time''
$\lambda\in\Sigma_\eps^1$.

Both the matrix models obtained by dimensional reductions of
$\CN=4$ supersymmetric SDYM theory and hCS theory are
(classically) equivalent. This follows from the dimensional
reduction of the formul\ae{} \eqref{fieldexpansion} defining the
Penrose-Ward transform. The reduced superfield expansion is fixed
by the geometry of $\CP_\eps^{1|4}$ and reads explicitly as
\begin{subequations}\label{fieldexpansionred}
\begin{eqnarray}\label{expredAa}
\CX_\alpha^+&=&\lambda_+^\ald\,
A_{\alpha\ald}+\eta_i^+\chi^i_\alpha+
\gamma_+\,\tfrac{1}{2!}\,\eta^+_i\eta^+_j\,\hl^\ald_+\,
\phi_{\alpha \ald}^{ij}+\\
\nonumber
&&+\gamma_+^2\,\,\tfrac{1}{3!}\,\eta^+_i\eta^+_j\eta^+_k\,\hl_+^\ald\,
\hl_+^\bed\,
\tilde{\chi}^{ijk}_{\alpha\ald\bed}+\gamma_+^3\,\tfrac{1}{4!}\,
\eta^+_i\eta^+_j\eta^+_k\eta^+_l\,
\hl_+^\ald\,\hl_+^\bed\,\hl_+^{\dot{\gamma}}\,
G^{ijkl}_{\alpha\ald\bed\dot{\gamma}}~,\\
\label{expredAl}
\CA_{\bl_+}&=&\gamma_+^2\eta^+_i\eta^+_j\,\phi^{ij}-
\gamma_+^3\eta^+_i\eta^+_j\eta^+_k\,\hl_+^\ald\,
\tilde{\chi}^{ijk}_{\ald}
+\\
\nonumber &&+2\gamma_+^4\eta^+_i\eta^+_j\eta^+_k\eta^+_l\,
\hl_+^\ald\,\hl_+^\bed G^{ijkl}_{\ald\bed}~,
\end{eqnarray}
\end{subequations}
where all component fields are independent of $x\in\FR^4$. One can
substitute this expansion into the action \eqref{actionhCSred} and
after a subsequent integration over $\CP^{1|4}_\eps$, one obtains
the action \eqref{SDYMactionMM} up to a constant multiplier, which
is the volume\footnote{In the Kleinian case, this volume is
na\"ively infinite, but one can regularize it by utilizing a
suitable partition of unity.} of $\Sigma^1_\eps$.

\subsection{Noncommutative $\CN=4$ SDYM theory}

Noncommutative field theories have received much attention
recently, as they were found to arise in string theory in the
presence of D-branes and a constant NS $B$-field background
\cite{Seiberg:1999vs,Douglas:2001ba,Szabo:2001kg}.

There are two completely equivalent ways of introducing a
noncommutative deformation of classical field theory: a
star-product formulation and an operator formalism. In the first
approach, one simply deforms the ordinary product of classical
fields (or their components) to the noncommutative star product
which reads in spinor notation as
\begin{equation}
(f\star g)(x)\ :=\
f(x)\exp\left(\tfrac{\di}{2}\overleftarrow{\dpar_{\alpha\ald}}
\theta^{\alpha\ald\beta\bed}\overrightarrow{\dpar_{\beta\bed}}\right)g(x)
\end{equation}
with $\theta^{\alpha\ald\beta\bed}=-\theta^{\beta\bed\alpha\ald}$
and in particular
\begin{equation}
x^{\alpha\ald}\star x^{\beta\bed}-x^{\beta\bed}\star
x^{\alpha\ald}\ =\ \di \theta^{\alpha\ald\beta\bed}~.
\end{equation}
In the following, we restrict ourselves to the case of a self-dual
$(\kappa=1)$ or an anti-self-dual $(\kappa=-1)$ tensor
$\theta^{\alpha\ald\beta\bed}$ and choose coordinates such that
\begin{equation}
\theta^{1\ed 2\zd}\ =\ -\theta^{2\zd 1\ed}\ =\ -2\di\kappa\eps
\theta\eand \theta^{1\zd 2\ed}\ =\ -\theta^{2\ed 1\zd}\ =\ 2\di
\eps\theta~.
\end{equation}

The formulation of noncommutative $\CN=4$ SDYM theory on
$(\FR^4_\theta,g_\eps)$ is now achieved by replacing all products
in the action \eqref{SDYMaction} by star products. For example,
the noncommutative field strength will read
\begin{equation}
F_{\alpha\ald\beta\bed}\ =\ \dpar_{\alpha\ald}A_{\beta\bed}-\dpar_{\beta\bed}A_{\alpha\ald}+
A_{\alpha\ald}\star A_{\beta\bed}-A_{\beta\bed}\star
A_{\alpha\ald}~.
\end{equation}

For the matrix reformulation of our model, it is necessary to
switch to the operator formalism, which trades the star product
for operator-valued coordinates $\hat{x}^{\alpha\ald}$ satisfying
\begin{equation}\label{Heisenbergalgebra}
[\hat{x}^{\alpha\ald},\hat{x}^{\beta\bed}]\ =\ \di
\theta^{\alpha\ald\beta\bed}~.
\end{equation}
This defines the noncommutative space $\FR^4_\theta$ and on this
space, derivatives are inner derivations of the Heisenberg algebra
\eqref{Heisenbergalgebra}:
\begin{equation}
\begin{aligned}
\der{\hat{x}^{1\ed}} f&\ :=\
-\frac{1}{2\kappa\eps\theta}[\hat{x}^{2\zd},f]~,~~~
&\der{\hat{x}^{2\zd}} f&\ :=\
+\frac{1}{2\kappa\eps\theta}[\hat{x}^{1\ed},f]~,\\
\der{\hat{x}^{1\zd}} f&\ :=\
+\frac{1}{2\eps\theta}[\hat{x}^{2\ed},f]~,~~~
&\der{\hat{x}^{1\zd}} f&\ :=\
-\frac{1}{2\eps\theta}[\hat{x}^{1\zd},f]~.
\end{aligned}
\end{equation}
The obvious representation space for the algebra
\eqref{Heisenbergalgebra} is the two-oscillator Fock space $\CH$
which is created from a vacuum state $|0,0\rangle$. This vacuum
state is annihilated by the operators
\begin{equation}
\hat{a}_1\ =\
\di\left(\frac{1-\eps}{2}\hat{x}^{2\ed}+\frac{1+\eps}{2}\hat{x}^{1\zd}\right)\eand
\hat{a}_2\ =\
-\di\left(\frac{1-\kappa\eps}{2}\hat{x}^{2\zd}+\frac{1+\kappa\eps}{2}\hat{x}^{1\ed}\right)
\end{equation}
and all other states of $\CH$ are obtained by acting with the
corresponding creation operators on $|0,0\rangle$. Thus,
coordinates as well as fields are to be regarded as operators in
$\CH$.

Via the Moyal-Weyl map
\cite{Seiberg:1999vs,Douglas:2001ba,Szabo:2001kg}, any function
$\Phi(x)$ in the star-product formulation can be related to an
operator-valued function $\hat{\Phi}(\hat{x})$ acting in $\CH$.
This yields the operator equivalent of star multiplication and
integration
\begin{equation}
f\star g\ \mapsto\ \hat{f}\hat{g}\eand \int \dd^4 x f\ \mapsto\
(2\pi\theta)^2\tr_\CH \hat{f}~,
\end{equation}
where $\tr_\CH$ signifies the trace over the Fock space $\CH$.

We now have all the ingredients for defining noncommutative
$\CN=4$ super SDYM theory in the operator formalism. Starting
point is the analogue of the covariant derivatives which are given
by the formul\ae{}
\begin{equation}
\begin{aligned}\nonumber
\hat{X}_{1\dot{1}}&\ =\
-\frac{1}{2\kappa\eps\theta}\hat{x}^{2\dot{2}}\otimes
\unit_n+\hat{A}_{1\dot{1}}~,& \hat{X}_{2\dot{2}}&\ =\
\frac{1}{2\kappa\eps\theta}\hat{x}^{1\dot{1}}\otimes
\unit_n+\hat{A}_{2\dot{2}}~,\\
\hat{X}_{1\dot{2}}&\ =\
\frac{1}{2\eps\theta}\hat{x}^{2\dot{1}}\otimes
\unit_n+\hat{A}_{1\dot{2}}~,& \hat{X}_{2\dot{1}}&\ =\
-\frac{1}{2\eps\theta}\hat{x}^{1\dot{2}}\otimes
\unit_n+\hat{A}_{2\dot{1}}~.
\end{aligned}
\end{equation}
These operators act on the tensor product of the Fock space $\CH$
and the representation space of the Lie algebra of the gauge group
$\sU(n)$. The operator-valued field strength has then the form
\begin{equation}
\hat{F}_{\alpha\ald \beta\bed}\ =\
[\hat{X}_{\alpha\ald},\hat{X}_{\beta\bed}]+\di\theta_{\alpha\ald\beta\bed}\otimes
\unit_n~,
\end{equation}
where the tensor $\theta_{\alpha\ald\beta\bed}$ has components
\begin{equation}
\theta_{1\dot{1}2\dot{2}}\ =\ -\theta_{2\dot{2}1\dot{1}}\ =\
\di\frac{\kappa\eps}{2\theta}~,~~~ \theta_{1\dot{2}2\dot{1}}\ =\
-\theta_{2\dot{1}1\dot{2}}\ =\ -\di\frac{\eps}{2\theta}~,
\end{equation}
Recall that noncommutativity restricts the set of allowed gauge
groups and we therefore had to choose to work with $\sU(n)$
instead of $\sSU(n)$.

The action of noncommutative SDYM theory on
$(\FR^4_\theta,g_\eps)$ reads
\begin{equation}\label{SDYMactionNC}
\begin{aligned}
S^{\CN=4}_{\mathrm{ncSDYM}}\ =\
\tr_\CH\tr\Bigg(\Bigg.-\tfrac{1}{2}\eps^{\alpha\beta}&\hat{G}^{\ald\bed}
\left([\hat{X}_{\alpha\ald},\hat{X}_{\beta\bed}]+\di\theta_{\alpha\ald\beta\bed}\otimes
\unit_n\right)\\+
\tfrac{\eps}{2}\eps_{ijkl}\tilde{\hat{\chi}}^{\ald ijk}
[\hat{X}_{\alpha\ald},\hat{\chi}^{\alpha l}]&+
\tfrac{\eps}{2}\eps_{ijkl}
\hat{\phi}^{ij}[\hat{X}_{\alpha\ald},[\hat{X}^{\alpha\ald},\hat{\phi}^{kl}]]+
\eps_{ijkl}\hat{\phi}^{ij}\hat{\chi}^{\alpha
k}\hat{\chi}^l_{\alpha}\Bigg.\Bigg)~.
\end{aligned}
\end{equation}
For $\kappa=+1$, the term containing
$\theta_{\alpha\ald\beta\bed}$ vanishes after performing the index
sums. Note furthermore that in the limit of $n\rightarrow \infty$
for the gauge group $\sU(n)$, one can render the ordinary $\CN=4$
SDYM matrix model \eqref{SDYMactionMM} equivalent to
noncommutative $\CN=4$ SDYM theory defined by the action
\eqref{SDYMactionNC}. This is based on the fact that there is an
isomorphism of spaces $\FC^\infty\cong \CH$ and $\FC^n\otimes
\CH$.

\subsection{Noncommutative hCS theory}

The natural question to ask at this point is whether one can
translate the Penrose-Ward transform completely into the
noncommutative situation and therefore obtain a holomorphic
Chern-Simons theory on a noncommutative supertwistor space. For
the Penrose-Ward transform in the purely bosonic case, the answer
is positive (see e.g.\
\cite{Kapustin:2000ek,Takasaki:2000vs,Lechtenfeld:2001ie}).

In the supersymmetric case, simply by taking the correspondence
space to be the product space $(\FR^{4|8}_\theta,g_\eps)\times
\Sigma_\eps^1$ with the coordinate algebra
\eqref{Heisenbergalgebra} and unchanged algebra of Gra{\ss}mann
coordinates, we arrive together with the incidence relation in
\eqref{2.6} at noncommutative coordinates\footnote{Observe that
the coordinates on $\Sigma^1_\eps$ stay commutative.} on the
twistor space $\CP^{3|4}_{\eps,\theta}$ satisfying the relations
\begin{equation}\label{noncommz}
\begin{aligned}
{}[\hat{z}^1_\pm,\hat{z}^2_\pm]&\ =\ 2(\kappa-1)\eps
\lambda_\pm\theta~,&[\hat{\bar{z}}^1_\pm,\hat{\bar{z}}^2_\pm]&\ =\
-2(\kappa-1)\eps\bl_\pm \theta~,\\
{}[\hat{z}^1_+,\hat{\bar{z}}^1_+]&\ =\
2(\kappa\eps-\lambda_+\bl_+) \theta~,&
[\hat{z}^1_-,\hat{\bar{z}}^1_-]&\ =\ 2(\kappa\eps\lambda_-\bl_--1)
\theta~,\\
[\hat{z}^2_+,\hat{\bar{z}}^2_+]&\ =\
2(1-\eps\kappa\lambda_+\bl_+)\theta~,&[\hat{z}^2_-,\hat{\bar{z}}^2_-]&\
=\ 2(\lambda_-\bl_--\eps\kappa)\theta~,
\end{aligned}
\end{equation}
with all other commutators vanishing. Here, we clearly see the
advantage of choosing a self-dual deformation tensor $\kappa=+1$:
the first line in \eqref{noncommz} becomes trivial. We will
restrict our considerations to this case\footnote{Recall, however,
that the singularities of the moduli space of self-dual solutions
are not resolved when choosing a self-dual deformation tensor.} in
the following.

Thus, we see that the coordinates $z^\alpha$ and $\bz^\alpha$ are
turned into sections $\hat{z}^\alpha$ and $\hat{\bar{z}}^\alpha$
of the bundle $\CO(1)$ which are functions on $\CP^{1|4}_\eps$ and
take values in the the space of operators acting on the Fock space
$\CH$. The derivatives along the bosonic fibres of the fibration
$\CP^{3|4}_\eps\rightarrow \CP^{1|4}_\eps$ are turned into inner
derivatives of the algebra \eqref{noncommz}:
\begin{equation}
\der{\hat{\bz}^1_\pm}f\ =\ \frac{\eps}{2\theta}
\gamma_\pm\,[\hat{z}^1_\pm,f]~,~~~ \der{\hat{\bz}^2_\pm}f\ =\
\frac{1}{2\theta}\gamma_\pm\,[\hat{z}^2_\pm,f]~.
\end{equation}
Together with the identities \eqref{eq:2.21}, we can furthermore
derive
\begin{equation}\label{Vdef}
\hat{\bV}_1^\pm f\ =\
-\frac{\eps}{2\theta}\,[\hat{z}^2_\pm,f]\eand \hat{\bV}_2^\pm f\
=\ -\frac{\eps}{2\theta}\,[\hat{z}^1_\pm,f]~.
\end{equation}

The formul\ae{} \eqref{Vdef} allow us to define the
noncommutatively deformed version of the hCS action
\eqref{actionhCScpt}:
\begin{align}\label{actionhCSdef1}
S_{\mathrm{nchCS}}\ :=\ & \int_{\CY_\eps}\omega\wedge \tr_\CH \tr
\left\{
\left(\hat{\CA}_2\dparb\hat{\CA}_1-\hat{\CA}_1\dparb\hat{\CA}_2\right)+2\hat{\CA}^{0,1}_\Sigma\left[\hat{\CA}_1,\hat{\CA}_2\right]-\right.
\\\nonumber&\left.-\frac{\eps}{2\theta}\left(\hat{\CA}_1\left[\hat{z}^1,\hat{\CA}^{0,1}_\Sigma\right]-
\hat{\CA}^{0,1}_\Sigma\left[\hat{z}^1,\hat{\CA}_1\right]+\hat{\CA}^{0,1}_\Sigma\left[\hat{z}^2,\hat{\CA}_2\right]-
\hat{\CA}_2\left[\hat{z}^2,\hat{\CA}^{0,1}_\Sigma\right]\right)\right\}~,
\end{align}
where $\CY_\eps$ is again the chiral subspace of $\CP^{1|4}_\eps$
for which $\etab^i_\pm=0$, $\omega$ is the form defined in
\eqref{defomega} and $\tr_\CH$ and $\tr$ denote the traces over
the Fock space $\CH$ and the representation space of
$\agl(n,\FC)$, respectively. The hats indicate that the components
of the gauge potential $\hat{\CA}^{0,1}$ are now operators with
values in the Lie algebra $\agl(n,\FC)$.

We can simplify the above action by introducing the operators
\begin{equation}
\hat{\CX}^1_\pm\ =\ -\frac{\eps}{2\theta}\hat{z}^2_\pm\otimes
\unit_n+\hat{\CA}_1^\pm\eand \hat{\CX}^2_\pm\ =\
-\frac{\eps}{2\theta}\hat{z}^1_\pm\otimes \unit_n+\hat{\CA}_2^\pm~
\end{equation}
which yields
\begin{equation}\label{actionhCSdef2}
S_{\mathrm{nchCS}}\ =\ \int_{\CY_\eps}\omega\wedge\tr_\CH
\tr\eps^{\alpha\beta}\hat{\CX}_\alpha\left(\dparb\hat{\CX}_\beta+
[\hat{\CA}^{0,1}_\Sigma,\hat{\CX}_\beta]\right)~,
\end{equation}
where the $\hat{\CX}_\alpha$ take values in the bundle $\CO(1)$,
so that the above integral is indeed well defined. Note that in
the matrix model \eqref{actionhCSred}, we considered matrices
taking values in the Lie algebra $\agl(n,\FC)$, while the fields
$\hat{\CX}_\alpha$ and $\hat{\CA}^{0,1}_\Sigma$ in the model
\eqref{actionhCSdef2} take values in $\agl(n,\FC)\otimes
\sEnd(\CH)$ and can be represented by infinite dimensional
matrices.

\subsection{String field theory}

It is of interest to generalize the twistor correspondence to the
level of string field theory (SFT). This could be done using the
approaches \cite{Berkovits:2004tx} or \cite{Siegel:2004dj}.
Alternatively, one could concentrate on (an appropriate extension
of) SFT for $\CN=2$ string theory \cite{Neitzke:2004pf}. At tree
level, open $\CN=2$ strings are known to reduce to the SDYM model
in a Lorentz noninvariant gauge \cite{Ooguri:ff}; their SFT
formulation \cite{Berkovits:1995ab} is based on the $\CN=4$
topological string description \cite{Berkovits:ff}. The latter
contains twistors from the outset: The coordinate
$\lambda\in\CPP^1$, the linear system, and the classical solutions
with the help of twistor methods were all incorporated in $\CN=2$
open string theory \cite{Lechtenfeld:ff}. Since this theory
\cite{Berkovits:1995ab} generalizes the Wess-Zumino-Witten-type
model \cite{Nair:1991ab} for SDYM theory and thus describes only
self-dual gauge fields (having helicity +1), it is not Lorentz
invariant. Its maximally supersymmetric extension $\CN=4$ super
SDYM theory, however, does admit a Lorentz-invariant formulation
\cite{Siegel:1992za,Chalmers:1996rq}. This theory features pairs
of fields of opposite helicity. In \cite{Berkovits:1997pq}, it was
proposed to lift the corresponding Lagrangian to SFT and in
\cite{Lechtenfeld:2004cc}, the twistor description of this model
was given as a specialization of Witten's supertwistor SFT when
one allows the string to vibrate only in part of the supertwistor
space (not in $\CPP^1\embd \CP^{3|4}$). The form of the matrix
model action given by \eqref{actionhCSdef2} is identical to an
action of this cubic string field theory for open $\CN=2$ strings
\cite{Lechtenfeld:2004cc}. Let us comment on that point in more
detail.

First of all, recall the definition of cubic open string field
theory \cite{Witten:1986qs}. Take a $\RZ$-graded algebra $\frA$
with an associative product $\star$ and a derivative $Q$ with
$Q^2=0$ and $|Q \CA|=|\CA|+1$ for any $\CA\in\frA$. Assume
furthermore a map $\int:\frA\rightarrow \FC$ which gives
non-vanishing results only for elements of grading $3$ and
respects the grading, i.e.\ $\int \CA\star
\CB=(-1)^{|\CA||\CB|}\int \CB\star \CA$. The (formal) action of
cubic string field theory is then
\begin{equation}
S=\tfrac{1}{2}\int \left(\CA\star
Q\CA+\tfrac{2}{3}\CA\star\CA\star\CA\right)~.
\end{equation}
The action is invariant under the gauge transformations $\delta
\CA=Q\varphi-\varphi\star \CA+\CA\star \varphi$. One can easily
extend this action to allow for Chan-Paton factors by replacing
$\frA$ with $\frA\otimes \agl(n,\FC)$ and $\int$ with $\int\otimes
\tr$.

The physical interpretation of the above construction is the
following: $\CA$ is a ``string field'' encoding all possible
excitations of an open string. The operator $\star$ glues the
halves of two open strings to form a third one, and the operator
$\int$ folds an open string and glues its two halves together
\cite{Witten:1986qs}.

To qualify as a string field, $\CA$ is a functional of the
embedding map $\Phi$ from the string parameter space to the string
target space. For the case at hand, we take
\begin{equation}
\Phi:[0,\pi]\times G\ \rightarrow\  \CP^{3|4}_\eps~,
\end{equation}
where $\sigma\in[0,\pi]$ parameterizes the open string and $G\ni
v$ provides the appropriate set of Gra{\ss}mann variables on the
worldsheet. Expanding $\Phi(\sigma,v)=\phi(\sigma)+v\psi(\sigma)$,
this map embeds the $\CN=2$ spinning string into supertwistor
space. Next, we recollect
$\phi=(z^\alpha=x^{\alpha\ald}\lambda_\ald,\eta_i=\eta_i^\ald\lambda_\ald,\lambda,\bl)$
and allow the string to vibrate only in the $z^\alpha$-directions
but keep the $G$-even zero modes of $(\eta_i,\lambda,\bl)$, so
that the string field depends on
$\{z^{\alpha}(\sigma),\eta_i,\lambda,\bl;\psi^{\alpha\ddot{\alpha}}(\sigma)\}$
only \cite{Lechtenfeld:2004cc}. Note that with $\psi$ and $\eta$,
we have two types of fermionic fields present, since we are
implicitly working in the doubly supersymmetric description of
superstrings \cite{Sorokin:1999jx}, which we will briefly discuss
in section~4.2. Therefore, the two fermionic fields are linked via
a superembedding condition. We employ a suitable BRST operator
$Q=\bar{D}+\dparb$, where
$\bar{D}=\psi^{\alpha\ddot{1}}\lambda^\ald\dpar_{\sigma}x_{\alpha\ald}\in\CO(1)$
and $\dparb\in\CO(0)$ are type $(0,1)$ vector fields on the fibres
and the base of $\CP^{3|4}_\eps$, respectively, and split the
string field accordingly, $\CA=\CA_{\bar{D}}+\CA_{\dparb}$. With a
holomorphic integration measure on $\CP^{3|4}_\eps$, the
Chern-Simons action \eqref{actionhCS} projects to
\cite{Lechtenfeld:2004cc}
\begin{equation}\label{actionSFT}
S\ =\ \int_{\CY_\eps} \omega \wedge\langle \tr
(\CA_{\bar{D}}\star\dparb\CA_{\bar{D}}+2\CA_{\bar{D}}\star
\bar{D}\CA_{\dparb}+
2\CA_{\dparb}\star\CA_{\bar{D}}\star\CA_{\bar{D}}\rangle~.
\end{equation}
Note that the string fields $\CA_{\bar{D}}$ and $\CA_{\dparb}$ are
fermionic, i.e.\ they behave in the action as if they were forms
multiplied with the wedge product. Furthermore, the
above-mentioned $\RZ$-grading of all the ingredients of this
action has to be adjusted appropriately. Giving an expansion in
$\eta_i$ for these string fields similar to the one in
\eqref{expAa} and \eqref{expAl}, one recovers the super string
field theory proposed by Berkovits and Siegel
\cite{Berkovits:1997pq}. Its zero modes describe self-dual $\CN=4$
SDYM theory.

By identifying $Q+\CA_{Q}$ with $\hat{\CX}$, $\dparb$ with
$\dparb_{\bl}$ and $\CA_\dparb$ with $\hat{\CA}^{0,1}_\Sigma$ and
adjusting the $\RZ_2$-grading of the fields, one obtains the
action\footnote{Note that $\dparb Q+Q\dparb=0$.}
\eqref{actionhCSdef2} from \eqref{actionSFT}. Therefore, we can
e.g.\ translate solution generating techniques which are at hand
for our matrix model immediately to the string field theory
\eqref{actionSFT}.

\section{Identification with D-brane configurations}

This section is devoted to presenting an interpretation of our
matrix models in terms of D-brane configurations in superstring
theory. After briefly reviewing some aspects of ordinary D-branes
and super D-branes, we will relate the matrix models arising from
hCS theory to topological D-branes. It follows the interpretation
of the corresponding matrix model arising from SDYM theory as
D-branes within $\CN=2$ string theory. Then we will switch to
ten-dimensional $\CN=1$ superstrings and find a connection of our
matrix models with a supersymmetric version of the ADHM
construction. By adding a term to the matrix model action
\eqref{actionhCSred}, we can even map all ingredients of the
D-brane interpretation of the ADHM prescription from the moduli
space $\FR^{4|8}$ to the twistor space $\CP^{3|4}_\eps$.
Eventually, we comment on the matrix models obtained from
noncommutativity.

Note that we use different conventions in $\CN=1$ and $\CN=2$
critical superstring theories: The worldvolume of a D$p$-brane is
meant to have dimension $(1,p)$ and $(a,b)$ with $a+b=p$,
respectively.

\subsection{Review of ordinary D-branes within D-branes}

In type IIB superstring theory, the Ramond-Ramond sector contains
$i$-form fields $C_{(i)}$ for $i=0,2,4,6,8,10$, which couple
naturally to D-branes of spatial dimension $i-1$. Recall that
there are two different points of view for these D-branes. First,
one can understand a D$p$-brane as a $p$-dimensional hyperplane on
which open strings end. Second, a D$p$-brane is a soliton of type
IIB supergravity in ten dimensions.

A stack of $n$ D$p$-branes naturally comes with a rank $n$ vector
bundle $E$ over their $p+1$-dimensional worldvolume together with
a connection one-form $A$. This field arises from the Chan-Paton
factors attached to the ends of an open string. The equations
determining the dynamics of $A$ at low energies in a flat
background are just the $\CN=1$ super Yang-Mills equations with
gauge group $\sU(n)$, dimensionally reduced from ten to $p+1$
dimensions. The emerging Higgs fields determine the movement of
the D$p$-brane in its normal directions. On K\"ahler manifolds,
the BPS sector is given by (a supersymmetric extension of) the
Hermitean Yang-Mills equations\footnote{or ``generalized Hitchin
equations''}
\begin{equation}\label{hym}
F^{0,2}\ =\ 0 \ =\ F^{2,0} \eand k^{d-1}\wedge F\ =\ \gamma k^d~,
\end{equation}
which are also reduced appropriately from ten to $p+1$ dimensions,
see e.g. \cite{Iqbal:2003ds}. Here, $k$ is the K\"{a}hler form of the
target space and $\gamma$ is the slope of $E$, i.e.\ a constant
enconding information about the first Chern class of the vector
bundle $E$. These equations imply the (dimensionally reduced,
supersymmetric) Yang-Mills equations.

If we just consider the topological subsector of the theory, the
dynamics of the connection one-form $A$ is described by an
appropriate dimensional reduction of the holomorphic Chern-Simons
equations \cite{Witten:1992fb,Vafa:2001qf,Neitzke:2004ni}, which
are given by
\begin{equation}
F^{0,2}\ =\ 0 \ =\ F^{2,0}~.
\end{equation}
Thus, the dynamics of topological D-branes differs from the one of
their BPS-cousins only by the second equation in \eqref{hym},
which is a stability condition on the vector bundle $E$.

A bound state of a stack of D$p$-branes with a D($p$-4)-brane can
be described in two different ways. On the one hand, we can look
at this state from the perspective of the higher-dimensional
D$p$-brane. Here, we find that the D($p$-4) brane is described by
a gauge field strength $F$ on the bundle $E$ over the worldvolume
of the D$p$-brane with a nontrivial second Chern character
$ch_2(E)$. The instanton number (the number of D($p$-4) branes) is
given by the corresponding second Chern class. In particular, the
BPS bound state of a stack of D$3$-branes with a D(-1)-brane is
given by a self-dual field strength $F=\ast F$ on $E$ with
$-\frac{1}{8\pi^2}\int F\wedge F=1$. On the other hand, one can
adopt the point of view of the D($p$-4)-brane inside the
D$p$-brane and consider the dimensional reduction of the $\CN=1$
super Yang-Mills equations from ten dimensions to the worldvolume
of the D($p$-4)-branes. To complete the picture, one has to add
strings with one end on the D$p$-brane and the other one on the
D($p$-4)-branes. Furthermore, one has to take into account that
the presence of the D$p$-brane will halfen the number of
supersymmetries once more, usually to a chiral subsector. In the
case of the above example of D$3$- and D(-1)-branes, this will
give rise to the ADHM equations discussed later. The situation for
bound states of D$p$- and D($p$-2)-branes can be discussed
analogously, and a bound state of stacks of D3- and D1-branes will
-- from the perspective of the D1-branes -- yield the Nahm
equations.

\subsection{Super D-branes}

As both the target space $\CP^{3|4}_\eps$ of the topological
B-model and the corresponding moduli space $\FR^{4|8}$ are
supermanifolds, we are naturally led to consider D-branes which
have also fermionic worldvolume directions.

Recall that there are three approaches of embedding worldvolumes
into target spaces when Gra{\ss}mann directions are involved. First,
one has the Ramond-Neveu-Schwarz (RNS) formulation
\cite{Neveu:1971rx,Ramond:1971gb}, which maps a super worldvolume
to a bosonic target space. This approach only works for a spinning
particle and a spinning string; no spinning branes have been
constructed so far. However, this formulation allows for a
covariant quantization. Second, there is the Green-Schwarz (GS)
formulation \cite{Green:1983sg}, in which a bosonic worldvolume is
mapped to a target space which is a supermanifold. In this
approach, the well-known $\kappa$-symmetry appears as a local
worldvolume fermionic symmetry. Third, there is the
doubly-supersymmetric formulation (see \cite{Sorokin:1999jx} and
references therein), which unifies in some sense both the RNS and
GS approaches. In this formulation, an additional superembedding
condition is imposed, which reduces the worldvolume supersymmetry
to the $\kappa$-symmetry of the GS approach.

In the following, we will always work implicitly with the doubly
supersymmetric approach.

\subsection{Topological D-branes and the matrix models}

The interpretation of the matrix model \eqref{actionhCSred} is now
rather straightforward. For gauge group $\sGL(n,\FC)$, it
describes a stack of $n$ almost space-filling D$(1|4)$-branes,
whose fermionic dimensions only extend into the holomorphic
directions of the target space $\CP^{3|4}_{\eps}$. These D-branes
furthermore wrap a $\CPP^{1|4}_{x}\embd \CP^{3|4}_{\eps}$.

We can use the expansion
$\CX_\alpha=\CX_\alpha^0+\CX_\alpha^i\eta_i+\CX_\alpha^{ij}\eta_i\eta_j+\ldots$
on any patch of $\CPP^{1|4}$ to examine the equations of motion
\eqref{shCS1} more closely:
\begin{equation}
\begin{aligned}
{}[\CX_1^0,\CX_2^0]&\ =\ 0~,\\
{}[\CX_1^i,\CX_2^0]+[\CX_1^0,\CX_2^i]&\ =\ 0~,\\
\{\CX_1^i,\CX_2^j\}-\{\CX_1^j,\CX_2^i\}+[\CX_1^{ij},\CX_2^0]+[\CX_1^0,\CX_2^{ij}]&\ =\ 0~,\\
\ldots
\end{aligned}
\end{equation}
Clearly, the bodies $\CX_\alpha^0$ of the Higgs fields can be
diagonalized simultaneously, and the diagonal entries describe the
position of the D$(1|4)$-brane in the normal directions of the
ambient space $\CP^{3|4}_\eps$. In the fermionic directions, this
commutation condition is relaxed and thus, the D-branes can be
smeared out in these directions even in the classical case.

\subsection{Interpretation within $\CN=2$ string theory}

The critical $\CN=2$ string has a four-dimensional target space
and its open string effective field theory is self-dual Yang-Mills
theory (or its noncommutative deformation
\cite{Lechtenfeld:2000nm} in the presence of a B-field). It has
been argued \cite{Siegel:1992za} that, after extending the $\CN=2$
string effective action in a natural way to recover Lorentz
invariance, the effective field theory becomes the full $\CN=4$
supersymmetrically extended SDYM theory, and we will adopt this
point of view in the following.

Considering D-branes in this string theory is not as natural as in
ten-dimensional superstring theories since the NS sector is
connected to the R sector via the $\CN=2$ spectral flow, and it is
therefore sufficient to consider the purely NS part of the $\CN=2$
string. Nevertheless, one can confine the end-points of the open
strings in this theory to certain subspaces and impose Dirichlet
boundary conditions to obtain objects which we will call D-branes
in $\CN=2$ string theory. Altough the meaning of these objects has
not yet been completely established, there seem to be a number of
safe statements we can recollect. First of all, the effective
field theory of these D-branes is four-dimensional
(supersymmetric) SDYM theory reduced to the appropriate
worldvolume \cite{Martinec:1997cw,Gluck:2003pa}. The
four-dimensional SDYM equations are nothing but the Hermitean
Yang-Mills equations:
\begin{equation}
F^{2,0}\ =\ F^{0,2}\ =\ 0\eand k\wedge F\ =\ 0~,
\end{equation}
where $k$ is again the K\"{a}hler form of the background.
The Higgs fields arising in the reduction process describe again
fluctuations of the D-branes in their normal directions.

As is familiar from the topological models yielding hCS theory, we
can introduce A- and B-type boundary conditions for the D-branes
in $\CN=2$ critical string theory. For the target space
$\FR^{2,2}$, the A-type boundary conditions are compatible with
D-branes of worldvolume dimension (0,0), (0,2), (2,0) and (2,2)
only \cite{Junemann:2001sp,Gluck:2003pa}.

Thus, we find a first interpretation of our matrix model
\eqref{SDYMactionMM} in terms of a stack of $n$ D0- or
D$(0|8)$-branes in $\CN=2$ string theory, and the topological
D$(1|4)$-brane is the equivalent configuration in B-type topological
string theory.

As usual, turning on a $B$-field background will give rise to
noncommutative deformations of the ambient space, and therefore
the matrix model \eqref{actionhCSdef2} describes a stack of $n$
D4-branes in $\CN=2$ string theory within such a background.

The moduli superspaces $\FR^{4|8}_\theta$ and $\FR^{0|8}$ for both
the noncommutative and the ordinary matrix model can therefore be
seen as {\em chiral} D$(4|8)$- and D$(0|8)$-branes, respectively,
with $\CN=4$ self-dual Yang-Mills theory as the appropriate
(chiral) low energy effective field theory.

\subsection{ADHM equations and D-branes}

The ADHM algorithm \cite{Atiyah:1978ri} for constructing instanton
solutions has found a nice interpretation in the context of string
theory \cite{Douglas:1995bn}; see also \cite{Tong:2005un} and
\cite{Dorey:2002ik} for a helpful review. We will follow the
discussion of the latter reference and start from a configuration
of $k$ D5-branes bound to a stack of $n$ D9-branes, which -- upon
dimensional reduction -- will eventually yield a configuration of
$k$ D(-1)-branes inside a stack of $n$ D3-branes.

{}From the perspective of the D5-branes, the $\CN=2$ supersymmetry
of type IIB superstring theory is broken down to $\CN=(1,1)$ on
the six-dimensional worldvolume of the D5-branes, which are BPS.
The fields in the ten-dimensional Yang-Mills multiplet are
rearranged into an $\CN=2$ vector multiplet
$(\phi_a,A_{\alpha\ald},\chi_\alpha^i,\mub^\ald_i)$, where the
indices $i=1,\ldots,4$, $a=1,\ldots 6$ and $\alpha,\ald=1,2$ label
the representations of the Lorentz group $\sSO(5,1)\sim\sSU(4)$
and the R-symmetry group $\sSO(4)\sim\sSU(2)_L\times \sSU(2)_R$,
respectively. Thus, $\phi$ and $A$ denote bosons, while $\chi$ and
$\mub$ refer to fermionic fields. Note that the presence of the
D$9$-branes will further break supersymmetry down to $\CN=(0,1)$
and therefore the above multiplet splits into the vector multiplet
$(\phi_a,\mub^\ald_i)$ and the hypermultiplet
$(A_{\alpha\ald},\chi_\alpha^i)$. In the following, we will
discuss the field theory on the D5-branes in the language of
$\CN=(0,1)$ supersymmetry.

Let us now consider the vacuum moduli space of this theory which
is called the Higgs branch. This is the sector of the theory,
where the $D$-field, i.e.\ the auxiliary field for the $\CN=(0,1)$
vector multiplet, vanishes\footnote{This is often referred to as
the $D$-flatness condition.}. {}From the Yang-Mills part
describing the vector multiplet, we have the contribution
$4\pi^2\alpha'{}^2\int \dd^6x\tr_k\frac{1}{2}D^2_{\mu\nu}$, where
we also introduce the notation
$D_{\mu\nu}=\tr_2(\vec{\sigma}\sigmab_{\mu\nu})\cdot \vec{D}$. The
hypermultiplet leads to an additional contribution of $\int
\dd^6x\tr_k\di
\vec{D}\cdot\vec{\sigma}^\ald{}_\bed\bar{A}^{\alpha\bed}A_{\alpha\ald}$.
Note that we use a bar instead of the dagger to simplify notation.
However, this bar must not be confused with complex conjugation.

It remains to include the contributions from open strings having
one end on a D5-brane and the other one on a D9-brane. These
additional degrees of freedom form two hypermultiplets under
$\CN=(0,1)$ supersymmetry, which sit in the bifundamental
representation of $\sU(k)\times\sU(n)$ and its conjugate. We
denote them by $(w_\ald,\psi^i)$ and
$(\bar{w}^\ald,\bar{\psi}^i)$, where $w_\ald$ and $\bar{w}^\ald$
and $\psi^i$ and $\bar{\psi}^i$ denote four complex scalars and
eight Weyl spinors, respectively. The contribution to the
$D$-terms is similar to the hypermultiplet considered above: $\int
\dd^6x\tr_k\di \vec{D}\cdot\vec{\sigma}^\ald{}_\bed\bar{w}^\bed
w_\ald$.

Collecting all the contributions of the $D$-field to
the action and varying them yields the equations of motion
\begin{equation}\label{Dflatness}
\alpha'{}^2\vec{D}\ =\ \frac{\di}{16\pi^2}\vec{\sigma}^\ald{}_\bed(\bar{w}^\bed
w_\ald+\bar{A}^{\alpha\bed}A_{\alpha\ald})~.
\end{equation}
After performing the dimensional reduction of the D5-brane to a
D(-1)-brane, the condition that $\vec{D}$ vanishes is
equivalent to the ADHM constraints.

Spelling out all possible indices on our fields, we have
$A_{\alpha\ald pq}$ and $w_{up\ald}$, where $p,q=1,\ldots,k$
denote indices of the representation $\mathbf{k}$ of the gauge
group $\sU(k)$ while $u=1,\ldots,n$ belongs to the $\mathbf{n}$ of
$\sU(n)$. Let us introduce the new combinations of indices
$r=u+p\otimes\alpha=1,\ldots,n+2k$ together with the matrices
\begin{equation}\label{rearrangefields}
(a_{r q\ald})\ =\ \left(\begin{array}{c} w_{uq\ald} \\
A_{\alpha\ald pq}
\end{array}\right)~,~~~
(\bar{a}^{\ald r}_q)\ =\
\left(\bar{w}^\ald_{qu}~~A_{pq}^{\alpha\ald}
\right)\eand (b^\beta_{r q})\ =\ \left(\begin{array}{c} 0 \\
\delta_\alpha{}^\beta\delta_{pq}
\end{array}\right)~,
\end{equation}
which are of dimension $(n+2k)\times 2k$, $2k\times(n+2k)$ and
$(n+2k)\times 2k$, respectively. Now we are ready to define a
$(n+2k)\times 2k$ dimensional matrix, the zero-dimensional Dirac
operator of the ADHM construction, which reads
\begin{equation}
\Delta_{r p \ald}(x)\ =\ a_{r p \ald}+b^\alpha_{r p}
x_{\alpha\ald}~,
\end{equation}
and we put $\bar{\Delta}_p^{\ald r}:=(\Delta_{r p\ald})^*$.
Written in the new components \eqref{rearrangefields}, the ADHM
constraints amounting to the $D$-flatness condition read
$\vec{\sigma}^\ald{}_\bed(\bar{a}^\bed a_\ald)=0$, or, more
explicitly,
\begin{equation}\label{ADHMconstraints}
\bar{a}_\ald a_\bed+\bar{a}_\bed a_\ald\ =\ 0~,
\end{equation}
where we defined as usual
$\bar{a}_\ald=\eps_{\ald\bed}\bar{a}^\bed$. All further
conditions, which are sometimes also summarized under ADHM
constraints, are automatically satisfied due to our choice of
$b^\alpha_{r p}$ and the reality properties of our fields.

The kernel of the zero-dimensional Dirac operator is generally of
dimension $n$, as this is the difference between its numbers of
rows and columns. It is spanned by vectors which can be arranged
to a complex matrix $U_{r u}$ satisfying
\begin{equation}
\bar{\Delta}_p^{\ald r}U_{r u}\ =\ 0~.
\end{equation}
Upon demanding that the frame $U_{r u}$ is orthonormal, i.e.\ that
$\bar{U}^r_u U_{r v}=\delta_{uv}$, we can construct a self-dual
$\sSU(n)$-instanton configuration from
\begin{equation}
(\CCA_{\alpha\ald})_{uv}\ =\ \bar{U}^r_u\dpar_{\alpha\ald} U_{r
v}~.
\end{equation}
Usually, one furthermore introduces the auxiliary matrix $f$ via
\begin{equation}
f\ =\ 2(\bar{w}^\ald
w_\ald+(A_{\alpha\ald}+x_{\alpha\ald}\otimes\unit_{k})^2)^{-1}~,
\end{equation}
which fits in the factorization condition $\bar{\Delta}_p^{\ald
r}\Delta_{r q\bed}=\delta^\ald_\bed(f^{-1})_{pq}$. Note that the
latter condition is again equivalent to the ADHM constraints
\eqref{ADHMconstraints} arising from \eqref{Dflatness}. The matrix
$f$ allows for an easy computation of the field strength
\begin{equation}\label{fieldstrength}
\CCF_{\mu\nu}\ =\ 4\bar{U}b\sigma_{\mu\nu}f\bar{b}U
\end{equation}
and the instanton number
\begin{equation}
k\ =\ -\frac{1}{16\pi^2}\int \dd^4 x \tr_n \CCF^2_{\mu\nu}\ =\
\frac{1}{16\pi^2}\int \dd^4 x \square^2\tr_k\log f~.
\end{equation}
The self-duality of $\CCF_{\mu\nu}$ in
\eqref{fieldstrength} is evident from the self-duality property of
$\sigma_{\mu\nu}$.

\subsection{Super ADHM construction and super D-branes}

Recall that there is a formulation of the ordinary super
Yang-Mills equations and their self-dual truncations in terms of
superfields, which we already used e.g.\ in
\eqref{SDYMconstraints}. In the superformulation, the field
content and the equations of motion take the same shape as in the
ordinary formulation, but with all the fields being superfields.
Moreover, one can find an Euler operator, which easily shows the
equivalence of the superfield equations with the ordinary field
equations, see e.g.\ \cite{Harnad:1985bc,Devchand:1996gv}.

For the super ADHM construction, let us consider $k$
D$(5|8)$-branes inside $n$ D$(9|8)$-branes. To describe this
scenario, it is only natural to extend the fields arising from the
strings in this configuration to superfields on $\FC^{10|8}$ and
the appropriate subspaces. In particular, we extend the fields
$w_\ald$ and $A_{\alpha\ald}$ entering into the bosonic
$D$-flatness condition to superfields living on $\FC^{6|8}$.
However, since supersymmetry is broken down to four copies of
$\CN=1$ due to the presence of the two stacks of D-branes, these
superfields can only be linear in the Gra{\ss}mann variables. From the
discussion in \cite{Harnad:1985bc}, we can then state what the
superfield expansion should look like:
\begin{equation}\label{expansion}
w_\ald\ =\ \z{w}_\ald+\psi^i\eta_{i\ald}\eand
A_{\alpha\ald}\ =\ \z{A}_{\alpha\ald}+\chi^i_\alpha\eta_{i\ald}~.
\end{equation}

After following the above discussion, we arrive at the
$D$-flatness condition
\begin{equation}\label{Dflatness2}
\alpha'{}^2\vec{D}\ =\
\frac{\di}{16\pi^2}\vec{\sigma}^\ald{}_\bed(\bar{w}^\bed
w_\ald+\bar{A}^{\alpha\bed}A_{\alpha\ald})\ =\ 0~,
\end{equation}
where now all fields are true superfields. After performing the
dimensional reduction of the D$(9|8)$-D$(5|8)$-brane configuration
to one containing D$(3|8)$- and D(-$1|8$)-branes and arranging the
resulting field content according to \eqref{rearrangefields}, we
can construct the zero-dimensional super Dirac
operator\footnote{One should stress, that an extension of the
Dirac operator to higher orders in the Gra{\ss}mann variables is
inconsistent with the ADHM construction, as is easily seen from
its original motivation via monads. The same is suggested from the
supersymmetries present in our D-brane configuration.}
\begin{equation}\label{superDirac}
\Delta_{r i \ald}\ =\ a_{r i \ald}+b^\alpha_{r i}x^R_{\alpha\ald}\
=\ \z{a}_{r i \ald}+b^\alpha_{r i}x^R_{\alpha\ald}+c_{r
i}^j\eta_{j \ald}~,
\end{equation}
where $(x_R^{\alpha\ald}, \eta_i^\ald)$ are coordinates on the
(anti-)chiral superspace $\FC^{4|8}$. That is, from the point of
view of the full superspace $\FC^{4|16}$ with coordinates
$(x^{\alpha\ald},\theta^{i\alpha},\eta_i^\ald)$, we have
$x_R^{\alpha\ald}=x^{\alpha\ald}+\theta^{i\alpha}\eta_i^\ald$. The
super ADHM constraints \eqref{Dflatness2} were discussed in
\cite{Semikhatov:wj} for the first time; see also
\cite{Araki:2005jn} for a related recent discussion.

In components, these super constraints \eqref{ADHMconstraints} read
\begin{equation}
\z{\bar{a}}_\ald \z{a}_\bed+\z{\bar{a}}_\bed \z{a}_\ald\ =\ 0~,~~~
\z{\bar{a}}_\ald c_i-\bar{c}_i \z{a}_\ald\ =\ 0~,~~~ \bar{c}_i
c_j-\bar{c}_j c_i\ =\ 0~.
\end{equation}
The additional sign in the equations involving $c_i$ arises from
ordering and extracting the Gra{\ss}mann variables $\eta_i^\ald$ as
well as the definition $\overline{c_i\eta_i^\ald}=\eta_i^\ald
\bar{c}_i=-\bar{c}_i\eta_i^\ald$.

As proven in \cite{Semikhatov:wj}, this super ADHM construction
gives rise to solutions to the $\CN=4$ self-dual Yang-Mills
equations in the form of the super gauge potentials
\begin{equation}\label{supersolutions}
\CCA_{\alpha\ald}\ =\ \bar{U}\dpar_{\alpha\ald} U\eand
\CCA^i_\ald\ =\ \bar{U} D^i_\ald U~,
\end{equation}
where $U$ and $\bar{U}$ are again zero modes of $\bar{\Delta}$ and
$\Delta$, normalized according to $\bar{U}U=\unit$. That is, the
super gauge potentials in \eqref{supersolutions} satisfy the
self-duality equations \eqref{SDYMconstraints}.

The fact that solutions to the $\CN=4$ SDYM equations in general
do not satisfy the $\CN=4$ SYM equations does not spoil our
interpretation of such solutions as D(-$1|8$)-branes, since in our
picture, $\CN=4$ supersymmetry is broken down to four copies of
$\CN=1$ supersymmetry. Note furthermore that $\CN=4$ SYM theory
and $\CN=4$ SDYM theory can be seen as different weak coupling
limits of {\em one} underlying field theory \cite{Witten:2003nn}.

\subsection{The SDYM matrix model and the super ADHM construction}

While a solution to the $\CN=4$ SDYM equations with gauge group
$\sU(n)$ and second Chern number $c_2=k$ describes a bound state
of $k$ D(-$1|8$)-branes with $n$ D$(3|8)$-branes at low energies,
the SDYM matrix model obtained by a dimensional reduction of this
situation describes a bound state between $k+n$ D(-$1|8$)-branes.
This implies that there is only one type of strings, i.e.\ those
having both ends on the D(-$1|8$)-branes. In the ADHM
construction, one can simply account for this fact by eliminating
the bifundamental fields, i.e. by putting $w_{\ald}$ and $\psi^i$
to zero.

Hence, the remaining ADHM constraints read
\begin{equation}
\vec{\sigma}^\ald{}_\bed (\bar{A}^{\alpha\bed}A_{\alpha\ald})\ =\ 0~,
\end{equation}
and one can use the reality conditions to show that these
equations are equivalent to
\begin{equation}
\eps^{\alpha\beta}[A_{\alpha\ald},A_{\beta\bed}]\ =\ 0~.
\end{equation}
The expansion \eqref{expansion} yields
\begin{equation}
[A_{\alpha\ald},\chi^{\alpha i}]\ =\ 0~.
\end{equation}
Thus, we recover the matrix SDYM equations \eqref{SDYMeomMM} with
fields of higher R~charges put to zero. This is expected since fields
with more than one R-symmetry index appear beyond linear order in the
Gra{\ss}mann fields, and their presence would spoil the ADHM construction.

\subsection{Extension of the matrix model}

It is now conceivable that the D3-D(-1)-brane\footnote{For
simplicity, let us suppress the fermionic dimensions of the
D-branes in the following.} system explaining the ADHM
construction can be carried over to the supertwistor space
$\CP^{3|4}_\eps$. To this end, we take a D1-D5-brane system and analyze
it, either via open D5-D5 strings with excitations corresponding in
the holomorphic Chern-Simons theory to gauge configurations with
non-trivial second Chern character, or else by looking at the D1-D1 and
the D1-D5 strings. The latter point of view gives rise to a
holomorphic Chern-Simons analogue of the ADHM configuration, as we
will show in the following.

The action for the D1-D1 strings is evidently our hCS matrix model
\eqref{actionhCSred}. To incorporate the D1-D5 strings, we can use
an action proposed by Witten in \cite{Witten:2003nn}\footnote{In
fact, he uses this action to complement the hCS theory in such a
way that it gives rise to full Yang-Mills theory on the moduli
space. For this, he changes the parity of the fields $\alpha$ and
$\beta$ to be fermionic.}
\begin{equation}\label{extendedaction}
\int_{\CY_\eps} \omega\wedge \tr(\beta \dparb \alpha +\beta
\CA^{0,1}_\Sigma \alpha)~,
\end{equation}
where the fields $\alpha$ and $\beta$ take values in the line
bundles $\CO(1)$ and transform in the fundamental and
antifundamental representation of the gauge group $\sGL(n,\FC)$,
respectively.

The equations of motion of the total matrix model, whose action is the sum
of \eqref{actionhCSred} and \eqref{extendedaction}, are then modified to
\begin{equation}\label{exteoms}
\begin{aligned}
\dparb\CX_\alpha+[\CA^{0,1}_\Sigma,\CX_\alpha]&\ =\ 0~,\\
[\CX_1,\CX_2]+\alpha\beta&\ =\ 0~,\\
\dparb\alpha+\CA^{0,1}_\Sigma\alpha\ =\ 0\eand
\dparb\beta&+\beta\CA^{0,1}_\Sigma\ =\ 0~.
\end{aligned}
\end{equation}
Similarly to the Higgs fields $\CX_\alpha$ and the gauge potential
$\CA^{0,1}_\Sigma$, we can give a general field expansion for
$\beta$ and $\alpha=\bar{\beta}$:
\begin{equation}\label{abexpansion}
\begin{aligned}
\beta_+\ =\ & \lambda_+^\ald
w_\ald+\psi^i\eta^+_i+\gamma_+\tfrac{1}{2!}\eta^+_i\eta^+_j\hl^\ald_+\rho^{ij}_\ald+
\gamma^2_+\tfrac{1}{3!}\eta^+_i\eta^+_j\eta^+_k\hl^\ald_+\hl^\bed_+\sigma_{\ald\bed}^{ijk}\\&+\gamma^3_+
\tfrac{1}{4!}\eta^+_i\eta^+_j\eta^+_k\eta^+_l\hl^\ald_+\hl^\bed_+\hl^\gad_+\tau_{\ald\bed\gad}^{ijkl}~,\\
\alpha_+\ =\ &
\lambda^\ald_+\eps_{\ald\bed}\bar{w}^\bed_++\bar{\psi}^i\eta^+_i+\ldots~.
\end{aligned}
\end{equation}
Applying the equations of motion, one learns that the fields
beyond linear order in the Gra{\ss}mann variables are composite
fields:
\begin{equation}
\rho^{ij}_\ald\ =\ w_\ald\phi^{ij}~,~~~ \sigma_{\ald\bed}^{ijk}\
=\ \tfrac{1}{2}w_{(\ald}\tilde{\chi}_{\bed)}^{ijk}\eand
\tau_{\ald\bed\gad}^{ijkl}\ =\
\tfrac{1}{3}w_{(\ald}G_{\bed\gad)}^{ijkl}~.
\end{equation}

We intentionally denoted the zeroth order components of $\alpha_+$
and $\beta_+$ by $\lambda^+_\ald\bar{w}^\ald$ and $\lambda^\ald_+
w_\ald$, respectively, since the expansion \eqref{abexpansion}
together with the field equations \eqref{exteoms} are indeed the
(super) ADHM equations
\begin{equation}
\vec{\sigma}^\ald{}_\bed(\bar{w}^\bed
w_\ald+A^{\alpha\ald}A_{\alpha\bed})\ =\ 0~,
\end{equation}
which are equivalent to the condition that
$\bar{\Delta}\Delta=\unit_2\otimes f^{-1}$. Again, the components
beyond linear order in the Gra{\ss}mann fields have been put to zero
in the Higgs fields $\CX_\alpha$ and the gauge potential
$\CA^{0,1}_\Sigma$ (which automatically does the same for the
fields $\alpha$ and $\beta$).

This procedure seems at first slightly ad-hoc, but again it
becomes quite natural, when recalling that for the ADHM D-brane
configuration, supersymmetry is broken from $\CN=4$ to four times
$\CN=1$. Furthermore, the fields which are put to zero give rise
to the potential terms in the action, and thus, we can regard
putting these fields to zero as an additional ``$D$-flatness
condition'' arising on the topological string side.

With this additional constraint, our matrix model
\eqref{actionhCSred} together with the extension
\eqref{extendedaction} is equivalent to the ADHM equations.
Therefore, it is dual to holomorphic Chern-Simons
theory on the full supertwistor space $\CP^{3|4}_\eps$,
in the same sense in which the ADHM construction is dual to SDYM theory.

Summarizing, the D3-D(-1)-brane system can be mapped via an
extended Penrose-Ward-transform to a D5-D1-brane system in
topological string theory. The arising super SDYM theory on the
D3-brane corresponds to hCS theory on the D5-brane, while the
matrix model describing the effective action on the D(-1)-brane
corresponds to our hCS matrix model on a topological D1-brane. The
additional D3-D(-1) strings completing the picture from the
perspective of the D(-1)-brane can be directly translated into
additional D5-D1 strings on the topological side. The ADHM
equations can furthermore be obtained from an extension of the hCS
matrix model on the topological D1-brane with a restriction on the
field content.

\subsection{D-branes in a nontrivial $B$-field background}

Except for the remarks on the $\CN=2$ string, we have not yet
discussed the matrix model which we obtained from deforming the
moduli space $\FR^{4|8}$ to a noncommutative spacetime.

In general, noncommutativity is interpreted as the presence of a
Kalb-Ramond $B$-field background in string theory. Thus, solutions
to the noncommutative SDYM theory \eqref{SDYMactionNC} on
$\FR^{4|8}_{\theta}$ are D(-$1|8$)-branes bound to a stack of
space-filling D$(3|8)$-branes in the presence of a $B$-field
background. This distinguishes the commutative from the
noncommutative matrix model: The noncommutative matrix model is
now dual to the ADHM equations, instead of being embedded like the
commutative one.

The matrix model on holomorphic Chern-Simons theory describes
analogously a topological almost space-filling D$(5|4)$-brane in the
background of a $B$-field. Note that a noncommutative deformation
of the target space $\CP^{3|4}_\eps$ does not yield any
inconsistencies in the context of the topological B-model. Such
deformations have been studied e.g.\ in \cite{Kapustin:2003sg} and
\cite{Iqbal:2003ds}.

On the one hand, we found two pairs of matrix models, which are
dual to each other (as the ADHM equations are dual to the SDYM
equations). On the other hand, we expect both pairs to be directly
equivalent to one another in a certain limit, in which the rank of
the gauge group of the commutative matrix model tends to infinity.
The implications of this observation might reveal some further
interesting features.

\section{Dimensional reductions related to the Nahm equations}

After the discussion of the ADHM construction in the previous
section, one is led to try to also translate the D-brane
interpretation of the Nahm construction to some topological
B-model on a Calabi-Yau supermanifold. This is in fact possible,
but since the D-brane configuration is somewhat more involved, we
will refrain from presenting details. In the subsequent
discussion, we strongly rely on results from \cite{Popov:2005uv},
where further details complementing our rather condensed
presentation can be found. As in this reference, we will constrain
our considerations to real structures yielding Euclidean
signature, i.e.\ $\eps=-1$.

\subsection{The D-brane interpretation of the Nahm construction}

Before presenting its super extension, let us briefly recollect
the ordinary Nahm construction \cite{Nahm:1979yw} starting from
its D-brane interpretation \cite{Diaconescu:1996rk} and
\cite{Hashimoto:2005yy}; see also \cite{Tong:2005un}. For
simplicity, we restrict ourselves to the case of
$\sSU(2)$-monopoles, but a generalization of our discussion to
gauge groups of higher rank is possible and rather
straightforward.

We start in ten-dimensional type IIB superstring theory with a
pair of D3-branes extended in the directions $1,2,3$ and located
at $x^4=\pm 1$, $x^M=0$ for $M>4$. Consider now a bound state of
these D3-branes with $k$ D1-branes extending along the $x^4$-axis
and ending on the D3-branes. As in the case of the ADHM
construction, we can look at this configuration from two different
points of view.

{}From the perspective of the D3-branes, the effective field theory
on their worldvolume is $\CN=4$ super Yang-Mills theory. The
D1-branes bound to the D3-branes and ending on them impose a BPS
condition, which amounts to the Bogomolny equations in three
dimensions,
\begin{equation}\label{Bogomolny}
D_a\Phi\ =\ \tfrac{1}{2}\eps_{abc}F_{bc}~,
\end{equation}
where $a,b,c=1,2,3$. The ends of the D1-branes act as magnetic
charges in the worldvolume of the D3-branes. They can
therefore be understood as magnetic monopoles
\cite{Callan:1997kz}, whose field configuration $(\Phi,A_a)$
satisfies the Bogomolny equations. These monopoles are static
solutions of the underlying Born-Infeld action.

{}From the perspective of the D1-branes, the effective field theory
is first $\CN=(8,8)$ super Yang-Mills theory in two dimensions,
but supersymmetry is broken by the presence of the two D3-branes
to $\CN=(4,4)$. As before, one can write down the corresponding
$D$-terms \cite{Tong:2005un} and impose a $D$-flatness condition:
\begin{equation}
D=\derr{X^a}{x^4}+[A_4,X^a]-\tfrac{1}{2}\eps_{abc}[X^b,X^c]+ R\ =\
0~,
\end{equation}
where the $X^a$ are the scalar fields corresponding to the
directions in which the D3-branes extend. The $R$-term is
proportional to $\delta(x^4\pm1)$ and allow for the D1-branes to
end on the D3-branes. It is related to the so-called Nahm boundary
conditions, which we do not discuss. The theory we thus found is
simply self-dual Yang-Mills theory, reduced to one dimension. By
imposing ``temporal gauge'' $A_4=0$, we arrive at the Nahm
equations
\begin{equation}\label{Nahm}
\derr{X^a}{s}-\tfrac{1}{2}\eps_{abc}[X^b,X^c]\ =\
0~~~\mbox{for}~~~-1\ <\ s\ <\ 1~,
\end{equation}
where we substituted $s=x^4$. From solutions to these (integrable)
equations, we can construct the one-dimensional Dirac operator
\begin{equation}\label{onedDirac}
\Delta^{\ald\bed}\ =\ (\unit_2)^{\ald\bed}\otimes\der{s}
+\sigma^{(\ald\bed)}_a(x^a-X^a)~.
\end{equation}
The equations \eqref{Nahm} are, analogously to the ADHM equations,
the condition for $\bar{\Delta}\Delta$ to commute with the Pauli
matrices, or equivalently, to have an inverse $f$:
\begin{equation}
\bar{\Delta}\Delta\ =\ \unit_2\otimes f^{-1}~.
\end{equation}
The normalized zero modes $U$ of the Dirac operator $\bar{\Delta}$
satisifying
\begin{equation}
\bar{\Delta}(s)U\ =\ 0~,~~~\int_{-1}^1 \dd s~ \bar{U}(s) U(s)\ =\
\unit
\end{equation}
then give rise to solutions to the Bogomolny equations
\eqref{Bogomolny} via the definitions
\begin{equation}
\varPhi(x)\ =\ \int_{-1}^1 \dd s~ \bar{U}(s)\,s\, U(s)\eand
\CCA_a(x)\ =\ \int_{-1}^1 \dd s~ \bar{U}(s)\,\der{x^a} U(s)~.
\end{equation}
The verification of this statement is straightforward when using
the identity
\begin{equation}
U(s)\bar{U}(s')\ =\
\delta(s-s')-\overrightarrow{\Delta}(s)f(s,s')\overleftarrow{\Delta}(s')~.
\end{equation}
Note that all the fields considered above stem from D1-D1 strings.
The remaining D1-D3 strings are responsible for imposing the BPS
condition and the Nahm boundary conditions for the $X^a$ at $s=\pm 1$.

The superextension of the Nahm construction is obtained,
analogously to the superextension of the ADHM construction, by
extending the Dirac operator \eqref{onedDirac} according to
\begin{equation}\label{onedDiracs}
\Delta^{\ald\bed}\ =\ (\unit_2)^{\ald\bed}\otimes
\der{s}+\sigma^{(\ald\bed)}_a(x^a-X^a)+(\eta^{(\ald}_i\chi^{\bed)i})~.
\end{equation}
The fields $\chi^{\ald i}$ are Weyl spinors and arise from the
D1-D1 strings. (More explicitly, consider a bound state of
D7-D5-branes, which dimensionally reduces to our D3-D1-brane
system. The spinor $\chi^{\ald i}$ is the spinor $\chi^i_\alpha$
we encountered before when discussing the $\CN=(0,1)$
hypermultiplet on the D5-brane.)

In the following, we will present a mapping to a configuration of
topological D-branes, analogously to the one previously found for
the ADHM construction.

\subsection{The superspaces $\CQ^{3|4}$ and $\hat{\CQ}^{3|4}$}

We want to consider a holomorphic Chern-Simons theory which
describes magnetic mono\-poles and their superextensions. For
this, we start from the holomorphic vector bundle
\begin{equation}
\CQ^{3|4}\ =\ \CO(2)\oplus\CO(0)\oplus \FC^4\otimes \Pi\CO(1)
\end{equation}
of rank $2|4$ over the Riemann sphere $\CPP^1$. This bundle is
covered by two patches $\tilde{\CV}_\pm$ on which we have the
coordinates $\lambda_\pm=w_2^\pm$ on the base space and
$w_1^\pm,w_3^\pm$ in the bosonic fibres. On the overlap
$\tilde{\CV}_+\cap\tilde{\CV}_-$, we have thus\footnote{The
labelling of coordinates is chosen to become as consistent as
possible with \cite{Popov:2005uv}.}
\begin{equation}\label{coordsQ56}
w^1_+\ =\ (w^2_+)^2 w^1_-~,~~~ w^2_+\ =\ \frac{1}{w^2_-}~,~~~ w^3_+\ =\ w^3_-~.
\end{equation}
The coordinates on the fermionic fibres of $\CQ^{3|4}$ are the
same as the ones on $\CP^{3|4}$, i.e.\ we have $\eta_i^\pm$ with
$i=1,\ldots 4$, satisfying $\eta_i^+=\lambda_+\eta_i^-$ on
$\tilde{\CV}_+\cap\tilde{\CV}_-$. From the Chern classes of the
involved line bundles, we clearly see that $\CQ^{3|4}$ is a
Calabi-Yau supermanifold.

Note that holomorphic sections of the vector bundle $\CQ^{3|4}$
are parameterized by elements $(y^{(\ald\bed)},y^4,\eta^\ald_i)$
of the moduli space $\FC^{4|8}$ according to
\begin{equation}
w^1_\pm\ =\
y^{\ald\bed}\lambda_\ald^\pm\lambda_\bed^\pm~,~~~w^3_\pm\ =\
y^4~,~~~ \eta_i^\pm\ =\ \eta_i^\ald\lambda_\ald^\pm\ewith
\lambda_\pm\ =\ w^2_\pm~.
\end{equation}

Let us now deform and restrict the sections of $\CQ^{3|4}$ by
identifying $y^4$ with $-\gamma_\pm\lambda^\pm_\ald\hl^\pm_\bed
y^{\ald\bed}$, where the coordinates $\hl_\ald$ were defined in
\eqref{alllambdas}. We still have $w^3_+=w^3_-$ on the overlap
$\tilde{\CV}_+\cap\tilde{\CV}_-$, but $w^3$ no longer describes a
section of a holomorphic line bundle. It is rather a section of a
smooth line bundle, which we denote by $\hat{\CO}(0)$. This
deformation moreover reduces the moduli space from $\FC^{4|8}$ to
$\FC^{3|8}$. We will denote the resulting total bundle by
$\hat{\CQ}^{3|4}$.

\subsection{Field theories and dimensional reductions}

First, we impose a reality condition on $\hat{\CQ}^{3|4}$ which is
(for the bosonic coordinates) given by
\begin{equation}
\tau(w_\pm^1,w_\pm^2)\ =\
\left(-\frac{\bar{w}_\pm^1}{(\bar{w}_\pm^2)^2},
-\frac{1}{\bar{w}_\pm^2}\right) \eand\tau(w^3_\pm)\ =\
\bar{w}^3_\pm~
\end{equation}
and keep complex the coordinate $w_\pm^2$ on the base $\CPP^1$, as usual.
Then $w^1_\pm$ remains complex, but $w^3_\pm$ becomes
real. In the identification with the real moduli
$(x^1,x^3,x^4)\in\FR^3$, we find that
\begin{equation}
y^{\ed\ed}\ =\ -(x^3+\di x^4)\ =\ -\bar{y}^{\zd\zd}\eand
w^3_\pm\ =\ x^1\ =\ -y^{\ed\zd}~.
\end{equation}
Thus, the space $\hat{\CQ}^{3|4}$ reduces to a Cauchy-Riemann (CR)
manifold\footnote{Roughly speaking, a CR manifold is a complex
manifold with additional real directions.}, which we label by
$\hat{\CQ}^{3|4}_{-1}=\CK^{5|8}$. This space has been extensively
studied in \cite{Popov:2005uv}, and it was found there that a
partially holomorphic Chern-Simons theory obtained from a certain
natural integrable distribution on $\CK^{5|8}$ is equivalent to
the supersymmetric Bogomolny model on $\FR^3$. Furthermore, it is
evident that the complexification of this partially holomorphic
Chern-Simons theory is holomorphic Chern-Simons theory on our
space $\hat{\CQ}^{3|4}$. This theory describes holomorphic
structures $\dparb_\CA$ on a vector bundle $\CE$ over
$\hat{\CQ}^{3|4}$, i.e.\ a gauge potential $\CA^{0,1}$ satisfying
$\dparb\CA^{0,1}+\CA^{0,1}\wedge \CA^{0,1}=0$.

There are now three possibilities for (bosonic) dimensional
reductions
\begin{equation}
\hat{\CQ}^{3|4}\ =\ \CO(2)\oplus\hat{\CO}(0)\oplus \FC^4\otimes
\Pi\CO(1) \ \rightarrow\ \left\{\begin{array}{l}
\CP^{2|4}\ :=\ \CO(2)\oplus \FC^4\otimes \Pi\CO(1) \\
\hat{\CQ}^{2|4}\ :=\ \hat{\CO}(0)\oplus \FC^4\otimes \Pi\CO(1) \\
\CPP^{1|4}\ :=\ \FC^4\otimes \Pi\CO(1)
\end{array}\right.~,
\end{equation}
which we want to discuss next.

The dimensional reduction of the holomorphic Chern-Simons theory
to the space $\CP^{2|4}$ has also been studied in
\cite{Popov:2005uv}. It yields a holomorphic BF-theory
\cite{holBFtheory}, where the scalar $B$-field originates as the
component $\der{\bar{w}^3_\pm}\lrcorner \CA^{0,1}$ of the gauge
potential $\CA^{0,1}$ on $\CE\rightarrow \hat{\CQ}^{3|4}$. This
theory is also equivalent to the above-mentioned super Bogomolny
model on $\FR^3$. It is furthermore the effective theory on a
topological D3-brane and -- via a Penrose-Ward transform -- can be
mapped to static BPS gauge configurations on a stack of D3-branes
in type IIB superstring theory. These gauge configurations have
been shown to amount to BPS D1-branes being suspended between the
D3-branes and extending in their normal directions. Therefore, the
holomorphic BF-theory is the topological analogue of the D3-brane
point of view of the D3-D1-brane system.

{}From the above discussion, the field theory arising from the
reduction to $\hat{\CQ}^{2|4}$ is also evident. Note that
considering this space is equivalent to considering
$\hat{\CQ}^{3|4}$ with the additional restriction
$y^{\ed\ed}=y^{\zd\zd}=0$. Therefore, we reduced the super
Bogomolny model from $\FR^3$ to $\FR^1$, and we arrive at a
(partially) holomorphic BF-theory, which is equivalent to
self-dual Yang-Mills theory in one dimension. Since this theory
yields precisely the gauge-covariant Nahm equations, we conclude
that this is the D1-brane point of view of the D3-D1-brane system.

The last reduction proposed above is the one to $\CPP^{1|4}$. This
amounts to a reduction of the super Bogomolny model from $\FR^3$
to a point, i.e.\ SDYM theory in zero dimensions. Thus, we arrive
again at the matrix models \eqref{actionhCSred} and
\eqref{SDYMactionMM} discussed previously. It is interesting to
note that the matrix model cannot tell whether it originated from
the space $\CP^{3|4}$ or $\hat{\CQ}^{3|4}$.

\subsection{The Nahm construction from topological D-branes}

In the previous section, we saw that both the physical D3-branes
and the physical D1-branes correspond to topological D3-branes
wrapping either the space $\CP^{2|4}\subset \hat{\CQ}^{3|4}$ or
$\hat{\CQ}^{2|4}\subset \hat{\CQ}^{3|4}$. The bound system of
D3-D1-branes therefore corresponds to a bound system of
D3-D3-branes in the topological picture. The two D3-branes are
separated by the same distance\footnote{In our presentation of the
Nahm construction, we chose this distance to be $1-(-1)=2$.} as
the physical ones in the normal direction $\CN_{\CP^{2|4}}\cong
\CO(2)$ in $\hat{\CQ}^{3|4}$. It is important to stress, however,
that since supersymmetry is broken twice by the D1- and the
D3-branes, in the topological picture, we have to put to zero all
fields except for $(A_a,\Phi,\chi^i_\ald)$.

It remains to clarify the r{\^o}le of the Nahm boundary conditions.
In \cite{Diaconescu:1996rk}, this was done by considering
a D1-brane probe in a T-dualized configuration consisting of D7-
and D5-branes. This picture cannot be translated easily into
twistor space. It would be interesting to see explicitly what the
boundary conditions correspond to in the topological setup.
Furthermore, it could be enlightening to study the topological
analogue of the Myers effect, which creates a funnel at the point
where the physical D1-branes end on the physical D3-branes.
Particularly the core of this ``bion'' might reveal interesting
features in the topological theory.

\section{Conclusions and outlook}

In this paper, we presented various dimensional reductions of both
holomorphic Chern-Simons theory on the supertwistor space
$\CP^{3|4}$ and the corresponding supersymmetric self-dual
Yang-Mills theory on $\FR^{4|8}$. In particular, we constructed
two matrix models, one on $\CPP^{1|4}$ and one on $\FR^{0|8}$,
whose solution spaces are bijective up to gauge transformations.
We also defined similar matrix models by introducing
noncommutativity on the moduli space of sections of the vector
bundle $\CP^{3|4}$ and treating both the supersymmetric self-dual
Yang-Mills theory and the holomorphic Chern-Simons theory on the
thus obtained deformed total space of $\CP^{3|4}$ in the operator
formalism.

Altogether, we obtained two matrix models on $\CPP^{1|4}$, with
actions closely related to $\CN=2$ string field theory, and also
two matrix models on $\FR^{0|8}$.

We furthermore gave an interpretation of the matrix models in
terms of D-brane configurations within B-type topological string
theory. During this discussion, we established connections between
topological branes and physical D-branes of type IIB superstring
theory, whose worldvolume theory had been reduced by an additional
BPS condition due to the presence of a further physical brane. Let
us summarize the correspondences in the following table:
\begin{equation}
\begin{aligned}
\mbox{D$(5|4)$-branes in $\CP^{3|4}_\eps$} &\ \leftrightarrow\
\mbox{D$(3|8)$-branes in $\FR^{4|8}$}\\
\mbox{D$(3|4)$-branes wrapping $\CP^{2|4}_\eps$ in
$\CP^{3|4}_\eps$ or $\hat{\CQ}^{3|4}_\eps$} &\ \leftrightarrow\
\mbox{static D$(3|8)$-branes in $\FR^{4|8}$}\\
\mbox{D$(3|4)$-branes wrapping $\hat{\CQ}^{2|4}_\eps$ in
$\hat{\CQ}^{3|4}_\eps$} &\ \leftrightarrow\
\mbox{static D$(1|8)$-branes in $\FR^{4|8}$}\\
\mbox{D$(1|4)$-branes in $\CP^{3|4}_\eps$} &\ \leftrightarrow\
\mbox{D(-$1|8$)-branes in $\FR^{4|8}$}~.
\end{aligned}
\end{equation}
It should be stressed that the fermionic parts of all the branes
in $\CP^{3|4}_\eps$ and $\hat{\CQ}^{3|4}_\eps$ only extend into
holomorphic directions. It is straightforward to add to this list
the diagonal line bundle $\CD^{2|4}_\eps$, which is obtained from
$\CP^{3|4}_\eps$ by imposing a condition\footnote{The condition
for the Euclidean case is slightly different.} $z^1_\pm=z^2_\pm$
on the local sections:
\begin{equation}
\mbox{D$(3|4)$-branes wrapping $\CD^{2|4}_\eps$ in $\CP^{3|4}_\eps$}
\leftrightarrow \mbox{ D$(1|8)$-branes in $\FR^{4|8}$}~.
\end{equation}

We furthermore established topological analogues of the D-brane
configurations underlying both the super ADHM and the super Nahm
construction. Thus, we found a matrix model over $\CPP^{1|4}$ and
a holomorphic BF-theory on $\hat{\CQ}^{2|4}$ which are dual to
holomorphic Chern-Simons theory on $\CP^{3|4}$ and the holomorphic
BF-theory on $\CP^{2|4}$, respectively, in the same sense the ADHM
and the Nahm equations are dual to the self-dual Yang-Mills and
the Bogomolny equations, respectively.

{}From the results presented in this paper, there arise a number
of interesting questions for further research. First, one should
examine in more detail the topological D-brane configuration
yielding the Nahm equations. In particular, it is desirable to
obtain more results on the Myers effect and the core of the
``bion'' in the topological setting as already mentioned above.
Second, one could imagine to strengthen and extend the relations
between D-branes in type IIB superstring theory and the
topological D-branes in the B-model. In the latter theory, the
powerful framework of derived categories (see e.g.\
\cite{Aspinwall:2004jr}) might then be carried over in some form
to the full ten-dimensional string theory. Eventually, it might
also be interesting to look at the mirror of the presented
configurations in the topological A-model.

\section*{Acknowledgements}

We would like to express our gratitude to Alexander D.~Popov for
many discussions and useful comments on this work. Furthermore, we
would like to thank Emanuel Scheidegger, Sebastian Uhlmann, Robert
Wimmer and Martin Wolf for helpful discussions. This work was
partially supported by the DFG priority program (SPP 1096) in
string theory.


\begin{thebibliography}{99}

\bibitem{Witten:2003nn}
E.~Witten, {\em Perturbative gauge theory as a string theory in
twistor space,} Commun.\ Math.\ Phys.\  {\bf 252} (2004) 189
[hep-th/0312171].

\bibitem{Berkovits:2004hg}
N.~Berkovits, {\em An alternative string theory in twistor space
for $\CN=4$ super-Yang-Mills,} Phys.\ Rev.\ Lett.\  {\bf 93}
(2004) 011601 [hep-th/0402045].

\bibitem{webpage}
The web-pages of the ``London Mathematical Society Workshop on
Twistor String Theory,'' Oxford 10-14 January 2005, available
online at {\tt http://www.maths.ox.ac.uk/$\sim$lmason/Tws/} and
the QMUL workshop ``From twistors to amplitudes'' at {\tt
http://www.strings.ph.qmul.ac.uk/$\sim$andreas/FTTA/}.

\bibitem{Popov:2004nk}
A.~D.~Popov and M.~Wolf, {\em Topological B-model on weighted
projective spaces and self-dual models in four dimensions,} JHEP
{\bf 0409} (2004) 007 [hep-th/0406224].

\bibitem{Ahn:2004xs}
C.~H.~Ahn, {\em Mirror symmetry of Calabi-Yau supermanifolds,}
Mod.\ Phys.\ Lett.\ A {\bf 20} (2005) 407 [hep-th/0407009].

\bibitem{Saemann:2004tt}
C.~Saemann, {\em The topological B-model on fattened complex
manifolds and subsectors of $\CN=4$ self-dual Yang-Mills theory,}
JHEP {\bf 0501} (2005) 042 [hep-th/0410292].

\bibitem{Park:2004bw}
J.~Park and S.~J.~Rey, {\em Supertwistor orbifolds: Gauge theory
amplitudes and topological strings,} JHEP {\bf 0412} (2004) 017
[hep-th/0411123].

\bibitem{Giombi:2004xv}
M.~Kulaxizi and K.~Zoubos, {\em Marginal deformations of $\CN=4$
SYM from open / closed twistor strings,} hep-th/0410122;
S.~Giombi, M.~Kulaxizi, R.~Ricci, D.~Robles-Llana, D.~Trancanelli
and K.~Zoubos, {\em Orbifolding the twistor string,} Nucl.\ Phys.\
B {\bf 719} (2005) 234 [hep-th/0411171].

\bibitem{Wolf:2004hp}
M.~Wolf, {\em On hidden symmetries of a super gauge theory and
twistor string theory,} JHEP {\bf 0502} (2005) 018
[hep-th/0412163].

\bibitem{Chiou:2005jn}
D.~W.~Chiou, O.~J.~Ganor, Y.~P.~Hong, B.~S.~Kim and I.~Mitra, {\em
Massless and massive three dimensional super Yang-Mills theory and
mini-twistor string theory,} Phys.\ Rev.\ D {\bf 71} (2005) 125016
[hep-th/0502076].

\bibitem{Popov:2005uv}
A.~D.~Popov, C.~Saemann and M.~Wolf, {\em The topological B-model
on a mini-supertwistor space and supersymmetric Bogomolny monopole
equations,} JHEP {\bf 0510} (2005) 058 [hep-th/0505161].

\bibitem{Saemann:2005ji}
C.~Saemann, {\em On the mini-superambitwistor space and $\CN=8$
super Yang-Mills theory,} hep-th/0508137.

\bibitem{Lindstrom:2005uh}
U.~Lindstrom, M.~Rocek and R.~von Unge, {\em Ricci-flat
supertwistor spaces,} hep-th/0509211.

\bibitem{Bonelli:2005dc}
G.~Bonelli, L.~Bonora and A.~Ricco, {\em Conifold geometries,
topological strings and multi-matrix models,} Phys.\ Rev.\ D {\bf
72} (2005) 086001 [hep-th/0507224].

\bibitem{Lechtenfeld:2004cc}
O.~Lechtenfeld and A.~D.~Popov, {\em Supertwistors and cubic
string field theory for open $\CN=2$ strings,} Phys.\ Lett.\ B
{\bf 598} (2004) 113 [hep-th/0406179].

\bibitem{Wimmer:2005bz}
R.~Wimmer, {\em D0-D4 brane tachyon condensation to a BPS state
and its excitation spectrum in noncommutative super Yang-Mills
theory,} JHEP {\bf 0505} (2005) 022 [hep-th/0502158].

\bibitem{Lechtenfeld:ff2}
Z.~Horvath, O.~Lechtenfeld and M.~Wolf, {\em Noncommutative
instantons via dressing and splitting approaches,} JHEP {\bf 0212}
(2002) 060 [hep-th/0211041];
O.~Lechtenfeld and A.~D.~Popov, {\em Noncommutative multi-solitons
in 2+1 dimensions,} JHEP {\bf 0111} (2001) 040 [hep-th/0106213];
{\em Scattering of noncommutative solitons in 2+1 dimensions,}
Phys.\ Lett.\ B {\bf 523} (2001) 178 [hep-th/0108118];
M.~Wolf, {\em Soliton antisoliton scattering configurations in a
noncommutative sigma model in 2+1 dimensions,} JHEP {\bf 0206}
(2002) 055 [hep-th/0204185];
M.~Ihl and S.~Uhlmann, {\em Noncommutative extended waves and
soliton-like configurations in $\CN=2$ string theory,} Int.\ J.\
Mod.\ Phys.\ A {\bf 18} (2003) 4889 [hep-th/0211263];
O.~Lechtenfeld, A.~D.~Popov and B.~Spendig, {\em Noncommutative
solitons in open $\CN=2$ string theory,} JHEP {\bf 0106} (2001)
011 [hep-th/0103196];
M.~T.~Grisaru, L.~Mazzanti, S.~Penati and L.~Tamassia, {\em Some
properties of the integrable noncommutative sine-Gordon system,}
JHEP {\bf 0404} (2004) 057 [hep-th/0310214];
O.~Lechtenfeld, L.~Mazzanti, S.~Penati, A.~D.~Popov and
L.~Tamassia, {\em Integrable noncommutative sine-Gordon model,}
Nucl.\ Phys.\ B {\bf 705} (2005) 477 [hep-th/0406065];
A.~V.~Domrin, O.~Lechtenfeld and S.~Petersen, {\em Sigma-model
solitons in the noncommutative plane: Construction and stability
analysis,} JHEP {\bf 0503} (2005) 045 [hep-th/0412001].

\bibitem{Popov:2004rb}
A.~D.~Popov and C.~Saemann, {\em On supertwistors, the
Penrose-Ward transform and $\CN=4$ super Yang-Mills theory,}
hep-th/0405123.

\bibitem{Harnad:1984vk}
J.~P.~Harnad, J.~Hurtubise, M.~Legar{\'e} and S.~Shnider, {\em
Constraint equations and field equations in supersymmetric $\CN=3$
Yang-Mills theory,} Nucl.\ Phys.\ B {\bf 256} (1985) 609.

\bibitem{Devchand:1996gv}
C.~Devchand and V.~Ogievetsky, {\em Interacting fields of
arbitrary spin and N$>$4 supersymmetric  self-dual Yang-Mills
equations,} Nucl.\ Phys.\ B {\bf 481} (1996) 188 [hep-th/9606027].

\bibitem{Siegel:1992za}
W.~Siegel, {\em The $\CN=2$ (4) string is selfdual $\CN=4$
Yang-Mills,} Phys.\ Rev.\ D {\bf 46} (1992) R3235
[hep-th/9205075].

\bibitem{Ishibashi:1996xs}
N.~Ishibashi, H.~Kawai, Y.~Kitazawa and A.~Tsuchiya, {\em A
large-N reduced model as superstring,} Nucl.\ Phys.\ B {\bf 498}
(1997) 467 [hep-th/9612115].

\bibitem{Seiberg:1999vs}
N.~Seiberg and E.~Witten, {\em String theory and noncommutative
geometry,} JHEP {\bf 9909} (1999) 032 [hep-th/9908142].

\bibitem{Douglas:2001ba}
M.~R.~Douglas and N.~A.~Nekrasov, {\em Noncommutative field
theory,} Rev.\ Mod.\ Phys.\  {\bf 73} (2001) 977 [hep-th/0106048].

\bibitem{Szabo:2001kg}
R.~J.~Szabo, {\em Quantum field theory on noncommutative spaces,}
Phys.\ Rept.\  {\bf 378} (2003) 207 [hep-th/0109162].

\bibitem{Kapustin:2000ek}
A.~Kapustin, A.~Kuznetsov and D.~Orlov, {\em Noncommutative
instantons and twistor transform,} Commun.\ Math.\ Phys.\  {\bf
221} (2001) 385 [hep-th/0002193].

\bibitem{Takasaki:2000vs}
K.~Takasaki, {\em Anti-self-dual Yang-Mills equations on
noncommutative spacetime,} J.\ Geom.\ Phys.\  {\bf 37} (2001) 291
[hep-th/0005194].

\bibitem{Lechtenfeld:2001ie}
O.~Lechtenfeld and A.~D.~Popov, {\em Noncommutative 't Hooft
instantons,} JHEP {\bf 0203} (2002) 040 [hep-th/0109209];
{\em Noncommutative monopoles and Riemann-Hilbert problems,} JHEP
{\bf 0401} (2004) 069 [hep-th/0306263].

\bibitem{Berkovits:2004tx}
N.~Berkovits and L.~Motl, {\em Cubic twistorial string field
theory,} JHEP {\bf 0404} (2004) 056 [hep-th/0403187].

\bibitem{Siegel:2004dj}
W.~Siegel, {\em Untwisting the twistor superstring,}
hep-th/0404255.

\bibitem{Neitzke:2004pf}
A.~Neitzke and C.~Vafa, {\em $\CN=2$ strings and the twistorial
Calabi-Yau,} hep-th/0402128.

\bibitem{Ooguri:ff}
H.~Ooguri and C.~Vafa, {\em Selfduality and $\CN=2$ string magic,}
Mod.\ Phys.\ Lett.\ A {\bf 5} (1990) 1389;
{\em Geometry of $\CN=2$ strings,} Nucl.\ Phys.\ B {\bf 361}
(1991) 469;
N.~Marcus, {\em A tour through $\CN=2$ strings,} hep-th/9211059;
O.~Lechtenfeld, {\em Mathematics and physics of $\CN=2$ strings,}
hep-th/9912281.

\bibitem{Berkovits:1995ab}
N.~Berkovits, {\em Super-Poincar{\'e} invariant superstring field
theory,} Nucl.\ Phys.\ B {\bf 450} (1995) 90 [Erratum-ibid.\ B
{\bf 459} (1996) 439] [hep-th/9503099].

\bibitem{Berkovits:ff}
N.~Berkovits and C.~Vafa, {\em $\CN=4$ topological strings,}
Nucl.\ Phys.\ B {\bf 433} (1995) 123 [hep-th/9407190];
H.~Ooguri and C.~Vafa, {\em All loop $\CN=2$ string amplitudes,}
Nucl.\ Phys.\ B {\bf 451}, 121 (1995) [hep-th/9505183].

\bibitem{Lechtenfeld:ff}
O.~Lechtenfeld and A.~D.~Popov, {\em On the integrability of
covariant field theory for open $\CN=2$ strings,} Phys.\ Lett.\ B
{\bf 494} (2000) 148 [hep-th/0009144];
O.~Lechtenfeld, A.~D.~Popov and S.~Uhlmann, {\em Exact solutions
of Berkovits' string field theory,} Nucl.\ Phys.\ B {\bf 637}
(2002) 119 [hep-th/0204155];
A.~Kling, O.~Lechtenfeld, A.~D.~Popov and S.~Uhlmann, {\em On
nonperturbative solutions of superstring field theory,} Phys.\
Lett.\ B {\bf 551} (2003) 193 [hep-th/0209186];
{\em Solving string field equations: New uses for old tools,}
Fortsch.\ Phys.\ {\bf 51} (2003) 775 [hep-th/0212335];
A.~Kling and S.~Uhlmann, {\em String field theory vertices for
fermions of integral weight,} JHEP {\bf 0307} (2003) 061
[hep-th/0306254];
M.~Ihl, A.~Kling and S.~Uhlmann, {\em String field theory
projectors for fermions of integral weight,} JHEP {\bf 0403}
(2004) 002 [hep-th/0312314].

\bibitem{Nair:1991ab}
V.~P.~Nair and J.~Schiff, {\em Kahler Chern-Simons theory and
symmetries of antiselfdual gauge fields,} Nucl.\ Phys.\ B {\bf
371}, 329 (1992).

\bibitem{Chalmers:1996rq}
G.~Chalmers and W.~Siegel, {\em The self-dual sector of {QCD}
amplitudes,} Phys.\ Rev.\ D {\bf 54} (1996) 7628 [hep-th/9606061].

\bibitem{Berkovits:1997pq}
N.~Berkovits and W.~Siegel, {\em Covariant field theory for
self-dual strings,} Nucl.\ Phys.\ B {\bf 505} (1997) 139
[hep-th/9703154].

\bibitem{Witten:1986qs}
E.~Witten, {\em Interacting field theory of open superstrings,}
Nucl.\ Phys.\ B {\bf 276} (1986) 291.

\bibitem{Sorokin:1999jx}
D.~P.~Sorokin, {\em Superbranes and superembeddings,} Phys.\
Rept.\ {\bf 329} (2000) 1 [hep-th/9906142].

\bibitem{Iqbal:2003ds}
A.~Iqbal, N.~Nekrasov, A.~Okounkov and C.~Vafa, {\em Quantum foam
and topological strings,} hep-th/0312022;
N.~Nekrasov, H.~Ooguri and C.~Vafa, {\em S-duality and topological
strings,} JHEP {\bf 0410} (2004) 009 [hep-th/0403167].

\bibitem{Witten:1992fb}
E.~Witten, {\em Chern-Simons gauge theory as a string theory,}
Prog.\ Math.\  {\bf 133} (1995) 637 [hep-th/9207094].

\bibitem{Vafa:2001qf}
C.~Vafa, {\em Brane/anti-brane systems and U(N$|$M) supergroup,}
hep-th/0101218.

\bibitem{Neitzke:2004ni}
A.~Neitzke and C.~Vafa, {\em Topological strings and their
physical applications,} hep-th/0410178.

\bibitem{Neveu:1971rx}
A.~Neveu and J.~H.~Schwarz, {\em Factorizable dual model of
pions,} Nucl.\ Phys.\ B {\bf 31} (1971) 86.

\bibitem{Ramond:1971gb}
P.~Ramond, {\em Dual theory for free fermions,} Phys.\ Rev.\ D
{\bf 3} (1971) 2415.

\bibitem{Green:1983sg}
M.~B.~Green and J.~H.~Schwarz, {\em Properties of the covariant
formulation of superstring theories,} Nucl.\ Phys.\ B {\bf 243}
(1984) 285.

\bibitem{Lechtenfeld:2000nm}
O.~Lechtenfeld, A.~D.~Popov and B.~Spendig, {\em Open $\CN=2$
strings in a B-field background and noncommutative self-dual
Yang-Mills,} Phys.\ Lett.\ B {\bf 507}, 317 (2001)
[hep-th/0012200].

\bibitem{Martinec:1997cw}
E.~J.~Martinec, {\em M-theory and $\CN=2$ strings,}
hep-th/9710122.

\bibitem{Gluck:2003pa}
D.~Gl\"{u}ck, Y.~Oz and T.~Sakai, {\em D-branes in $\CN=2$ strings,}
JHEP {\bf 0308} (2003) 055 [hep-th/0306112].

\bibitem{Junemann:2001sp}
K.~J\"{u}nemann and B.~Spendig, {\em D-brane scattering of $\CN=2$
strings,} Phys.\ Lett.\ B {\bf 520} (2001) 163 [hep-th/0108069].

\bibitem{Atiyah:1978ri}
M.~F.~Atiyah, N.~J.~Hitchin, V.~G.~Drinfeld and Y.~I.~Manin, {\em
Construction of instantons,} Phys.\ Lett.\ A {\bf 65} (1978) 185.

\bibitem{Douglas:1995bn}
E.~Witten, {\em Bound states of strings and p-branes,} Nucl.\
Phys.\ B {\bf 460} (1996) 335 [hep-th/9510135];
M.~R.~Douglas, {\em Branes within branes,} hep-th/9512077;
{\em Gauge fields and D-branes,} J.\ Geom.\ Phys.\ {\bf 28} (1998)
255 [hep-th/9604198].

\bibitem{Tong:2005un}
D.~Tong, {\em TASI lectures on solitons,} hep-th/0509216.

\bibitem{Dorey:2002ik}
N.~Dorey, T.~J.~Hollowood, V.~V.~Khoze and M.~P.~Mattis, {\em The
calculus of many instantons,} Phys.\ Rept.\  {\bf 371} (2002) 231
[hep-th/0206063].

\bibitem{Harnad:1985bc}
J.~P.~Harnad and S.~Shnider, {\em Constraints and field equations
for ten-dimensional super Yang-Mills theory,} Commun.\ Math.\
Phys.\ {\bf 106} (1986) 183.

\bibitem{Semikhatov:wj}
A.~M.~Semikhatov, {\em Supersymmetric instanton,} JETP Lett.\ {\bf
35} (1982) 560;
I.~V.~Volovich, {\em Superselfduality for supersymmetric
Yang-Mills theory,} Phys.\ Lett.\ B {\bf 123}, 329 (1983).

\bibitem{Araki:2005jn}
T.~Araki, T.~Takashima and S.~Watamura, {\em On a superfield
extension of the ADHM construction and $\CN=1$ super instantons,}
JHEP {\bf 0508} (2005) 065 [hep-th/0506112].

\bibitem{Kapustin:2003sg}
A.~Kapustin, {\em Topological strings on noncommutative
manifolds,} Int.\ J.\ Geom.\ Meth.\ Mod.\ Phys.\  {\bf 1} (2004)
49 [hep-th/0310057].

\bibitem{Nahm:1979yw}
W.~Nahm, {\em A simple formalism for the BPS monopole,} Phys.\
Lett.\ B {\bf 90} (1980) 413.

\bibitem{Diaconescu:1996rk}
D.~E.~Diaconescu, {\em D-branes, monopoles and Nahm equations,}
Nucl.\ Phys.\ B {\bf 503} (1997) 220 [hep-th/9608163].

\bibitem{Hashimoto:2005yy}
K.~Hashimoto and S.~Terashima, {\em Stringy derivation of Nahm
construction of monopoles,} JHEP {\bf 0509} (2005) 055
[hep-th/0507078].

\bibitem{Callan:1997kz}
C.~G.~Callan and J.~M.~Maldacena, {\em Brane dynamics from the
Born-Infeld action,} Nucl.\ Phys.\ B {\bf 513} (1998) 198
[hep-th/9708147].

\bibitem{holBFtheory}
A.~D.~Popov, {\em Holomorphic analogs of topological gauge
theories,} Phys.\ Lett.\ B {\bf 473} (2000) 65 [hep-th/9909135];
T.~A.~Ivanova and A.~D.~Popov, {\em Dressing symmetries of
holomorphic BF theories,} J.\ Math.\ Phys.\  {\bf 41} (2000) 2604
[hep-th/0002120];
L.~Baulieu and A.~Tanzini, {\em Topological symmetry of forms,
$\CN=1$ supersymmetry and S-duality on special manifolds,}
hep-th/0412014.

\bibitem{Aspinwall:2004jr}
P.~S.~Aspinwall, {\em D-branes on Calabi-Yau manifolds,}
hep-th/0403166.

\end{thebibliography}
\end{document}